\documentclass[preprint,12pt]{elsarticle}

\usepackage{amssymb}
\usepackage{amsmath}

\usepackage{booktabs}

\usepackage{hyperref}

\usepackage{xcolor}
\usepackage{bm,mathtools}
\usepackage{algorithm}
\usepackage{algpseudocode}
\usepackage{subfig}
\usepackage{graphicx}
\usepackage{tikz}
\usetikzlibrary{arrows.meta,positioning,matrix,fit,calc}

\usepackage{tabularx}
\usepackage{makecell}

\newcommand{\RR}{\mathbb{R}}
\newcommand{\I}{\bm I}
\newcommand{\B}{\bm B}
\newcommand{\E}{\bm E}

\newcommand{\x}{\bm x}

\newcommand{\W}{\mathcal{W}}

\newcommand{\transpose}{\mathsf{T}}
\newcommand{\mat}[1]{\mathbf{#1}}
\newcommand{\optionaltikzinput}[1]{%
  \IfFileExists{#1}{%
    \input{#1}%
  }{%
    \fbox{%
      \begin{minipage}{0.78\linewidth}
        \centering
        \vspace{1.2em}
        Missing TikZ figure source\\[0.4em]
        \texttt{\detokenize{#1}}
        \vspace{1.2em}
      \end{minipage}%
    }%
  }%
}

\journal{Journal of Computational Physics}

\begin{document}

\begin{frontmatter}

\title{Mass Matrix Assembly on Tensor Cores for Implicit Particle-In-Cell Methods}

\author{Luca Pennati, Stefano Markidis}

\affiliation{organization={KTH Royal Institute of Technology},
            city={Stockholm},
            postcode={114 28}, 
            country={Sweden}}

\begin{abstract}
Matrix-multiply-accumulate (MMA) units, or \emph{tensor cores}, are now widespread across modern computing architectures. Yet, their use for particle-grid operators remains limited. In implicit particle methods, mass-matrix assembly is a reduction-dominated kernel in which weighted outer products of interpolation weights are accumulated over particle support. We show that this operation can be reformulated exactly, cell by cell, as a sequence of matrix products matched to hardware MMA tiles. The formulation is general with respect to interpolation order and hardware platform, and applies to both scalar mass matrices and the tensorial block mass matrix arising in implicit in the Energy-Conserving Semi-Implicit Method (ECSIM) for Particle-in-Cell simulations. We introduce particle batching and a support-group decomposition for higher-order shape functions whose stencil extends beyond a single cell, specialize the method to first- and second-order B-spline interpolation, and implement it on NVIDIA tensor cores. The resulting kernels achieve up to  $3\times$ over optimized conventional implementations and reduce end-to-end ECSIM runtime by $\sim15\%$.
\end{abstract}


\end{frontmatter}

\section{Introduction}
\label{sec:introduction}
Modern computing architectures increasingly own a substantial fraction of their computational throughput to hardware Matrix-Multiply-Accumulate (MMA) units, commonly referred to as \emph{tensor cores} or \emph{matrix engines}.  Originally introduced to accelerate machine-learning workloads to perform an MMA operation per clock cycle~\cite{jouppi2017_tpu,Micikevicius2017mma_mixedPrecision,Reuthe2020mma_ml_accelerators,jouppi2023tpuV4}, these units are now available across essentially all major accelerator and processor families. However, their effective use in scientific computing, still depends on the ability to recast an application kernel into a sequence of small dense matrix products with sufficient regularity to match the underlying hardware tiles. MMA units have found broader application across additional scientific domains. In particular, tensor cores have been employed in linear algebra~\cite{Higham2022mixed_precision} and linear solvers~\cite{haidar2019mixedPrecision}, especially in the context of low and mixed-precision algorithms. Other studies have investigated how to recast finite element methods~\cite{cui2024_tc_fem} and general stencil operations~\cite{Liu2022tensor_cores_stencil} into MMA-friendly forms. Tensor cores have also been successfully applied to domain-specific problems, including signal processing~\cite{Oostrum2025tc_signalProcessing}, molecular docking~\cite{schieffer2023_autodock_tc}, and quantum molecular dynamics~\cite{Finkelstein2021quantum_tc}. However, tensor core applicability to irregular particle-grid operators remains much less explored and it is the topic of this paper.

An important example of such an operator is the mass-matrix assembly arising in semi-implicit particle methods. In the Energy-Conserving Semi-Implicit Method (ECSIM)~\cite{lapenta2017ecsim} for Particle-In-Cel (PIC) plasma simulations, the mass matrix represents the linear response of the plasma to the electric field and enters the field solve as a grid-defined operator assembled from particle information. Its construction requires, for each particle, the accumulation over all pairs of support nodes of a weighted outer product of interpolation values, scaled in ECSIM by a particle-dependent response tensor. Although mathematically simple, this operation is computationally demanding. It is dominated by fine grained reductions and irregular scatter patterns that map poorly to conventional Single Instruction Multiple Data (SIMD) and Single Instruction Multiple Threads (SIMT) execution. In practice, it becomes one of the most expensive stages of the ECSIM cycle. The same algebraic structure also appears more broadly in mass-matrix-based particle-grid formulations, including scalar variants in related methods, most notably in the Material Point Method (MPM) for continuum mechanics~\cite{burgess1992_flip_mass_matrix,sulsky1994_mpm,sulsky1995_mpm}, where a scalar mass matrix couples grid-node momenta. 

In this work, we show that the mass-matrix assembly can be reformulated exactly in a form that is naturally matched to tensor cores. We express the weighted outer-product accumulation as a tensor contraction over particles and decompose that contraction cell by cell. In this formulation, the local assembly reduces to a sequence of batched matrix products whose inner dimension coincides with the contraction dimension of MMA tiles. This approach leads to a general mapping strategy, independent of interpolation order and kind of matrix engine, and applies to both the scalar mass matrix and the tensorial block structure arising in ECSIM. In this work, we focus on ECSIM as the primary and most demanding case study.

The main contributions of this work are as follows. First, we present a general reformulation of the mass matrix assembly as a tensor contraction that factors into a matrix product  $M_{ab} = A_{ak}\, B_{kb}$, valid for arbitrary spatial dimensions and interpolation orders, applicable to both scalar and tensorial mass matrices. Second, we introduce a formal particle-batching and support-group decomposition strategy that maps the cell-local contraction onto fixed-size MMA tiles, with particular attention to higher-order shape functions whose stencil extends beyond a single cell.
Third, we describe a reference implementation on NVIDIA tensor cores for first-order (CIC) and second-order (TSC) B-spline interpolation in three dimensions, with performance benchmarks against optimized conventional GPU kernels that demonstrate an end-to-end simulation speedup.\\

The paper is organized as follows. In Section~\ref{sec:preliminaries} we introduce the ECSIM mass matrix, reformulate its assembly as a tensor contraction, and recall the B-spline shape functions used in this work. We also describe the hardware tensor cores abstraction used in the paper. In Section~\ref{sec:methodology} we develop the general mass matrix assembly strategy, including the cell-local outer-product decomposition, particle batching, support-group decomposition and sparse stencil deposition. In Section~\ref{sec:results} we present numerical results for our implementation on NVIDIA tensor cores, including comparisons against conventional GPU kernels and end-to-end speedup in a production kinetic plasma simulation. Section~\ref{sec:discussion} discusses the implications and limitations of the proposed approach, and Section~\ref{sec:conclusion} is dedicated to conclusions.

\section{Preliminaries}
\label{sec:preliminaries}
In the Particle-In-Cell (PIC) method~\cite{birdsall1991_pic,hockney1988computer}, particles evolve in Lagrangian coordinates while field quantities live on a discrete Eulerian grid, and interpolation functions mediate all field–particle interactions.
In ECSIM~\cite{lapenta2017ecsim}, the plasma medium response is represented by the \emph{mass matrix} operator, defined on the grid, which takes the form of a $3\times 3$ tensor block for every pair of grid nodes within the support of a particle shape function, thus requiring the assembly of nine components per node pair. Its construction involves accumulating, for each particle, the outer product of the corresponding shape-function weight vector with itself, scaled by a particle-dependent coefficient tensor. This procedure is traditionally implemented as a particle-by-particle scatter loop with fine-grained reductions, a computational pattern that maps poorly onto conventional accelerated architectures. As a result, mass-matrix assembly fails to fully exploit the computational capabilities of contemporary hardware and becomes the most time-consuming stage of the PIC cycle.

\subsection{Mass matrix in the ECSIM PIC method}
\label{sec:ecsim}   

We briefly recall the origin of the mass matrix in the ECSIM formulation by Lapenta~\cite{lapenta2017ecsim}, the reader is referred to that work for a complete derivation.

Consider a plasma described by $N_s$ species, each represented by $N_p^{(s)}$ computational particles on a $d$-dimensional Cartesian grid~$\mathcal{G}$ with nodes indexed by $g$.
Each particle at position $\x_p$ interacts with the grid through a compactly supported \emph{shape} (or \emph{weight}) function $W(\x_p - \x_g)$. We write
\begin{equation}
  W_{pg} \equiv W(\x_p - \x_g),
  \label{eq_shape_function}
\end{equation}
and denote by $\mathcal{N}_p$ the compact support of particle $p$, i.e.\ the set of grid nodes with nonzero weights:
\begin{equation}
\mathcal N_p \coloneqq \{\, g \in \mathcal G \mid W_{pg} \neq 0 \,\}.
\label{eq_supportG}
\end{equation}
Figure~\ref{fig_particle_grid_coupling} illustrates the particle-grid coupling in the two-dimensional case, for a first order Cloud-In-Cell (CIC) interpolation.
\begin{figure}[t!]
  \centering
\begin{tikzpicture}[
    x=2.0cm,
    y=2.0cm,
    particle/.style={circle, fill=red!80!black, draw=white, line width=0.6pt, inner sep=2.8pt},
    ptwog/.style={-{Latex[length=2.2mm,width=1.6mm]}, line width=0.9pt, orange!85!black},
    annot/.style={-{Latex[length=2.0mm,width=1.5mm]}, line width=0.8pt, black!70},
    gridline/.style={line width=0.75pt, gray!55}
  ]
    \fill[blue!8] (0,0) rectangle (1,1);
    \draw[gridline] (-0.35,0) -- (1.35,0);
    \draw[gridline] (-0.35,1) -- (1.35,1);
    \draw[gridline] (0,-0.35) -- (0,1.35);
    \draw[gridline] (1,-0.35) -- (1,1.35);
    \draw[line width=1.2pt, blue!45!black] (0,0) rectangle (1,1);

    \foreach \x/\y/\name in {0/0/g0,1/0/g1,0/1/g2,1/1/g3} {
      \coordinate (\name) at (\x,\y);
      \fill[blue!70!black] (\x,\y) circle (2.2pt);
    }

    \node[particle] (p) at (0.52,0.48) {};
    \node[font=\small, red!80!black, anchor=south west] at (0.50,0.58) {$p$};

    \draw[ptwog] (p) to[bend right=12] (g0);
    \draw[ptwog] (p) to[bend left=12]  (g1);
    \draw[ptwog] (p) to[bend left=12]  (g2);
    \draw[ptwog] (p) to[bend right=12] (g3);
    \node[font=\small, orange!85!black] at (0.5,1.22) {$W_{gp}$};

    \draw[annot] (1.55,0.95) -- (1.03,0.97);
    \node[anchor=west, font=\small] at (1.60,0.95) {support nodes $g \in \mathcal{N}_p$};


\end{tikzpicture}
  \caption{Two-dimensional particle-grid coupling for the first-order (CIC) case. The particle~$p$ in the cell deposits to the four corner nodes in its support $\mathcal{N}_p$.}
  \label{fig_particle_grid_coupling}
\end{figure}
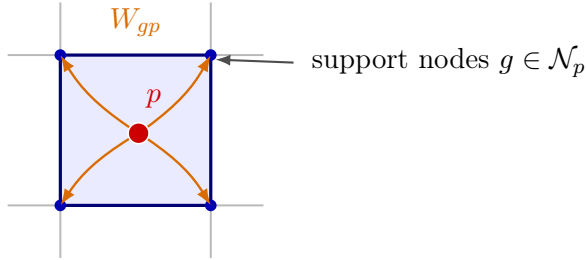\\

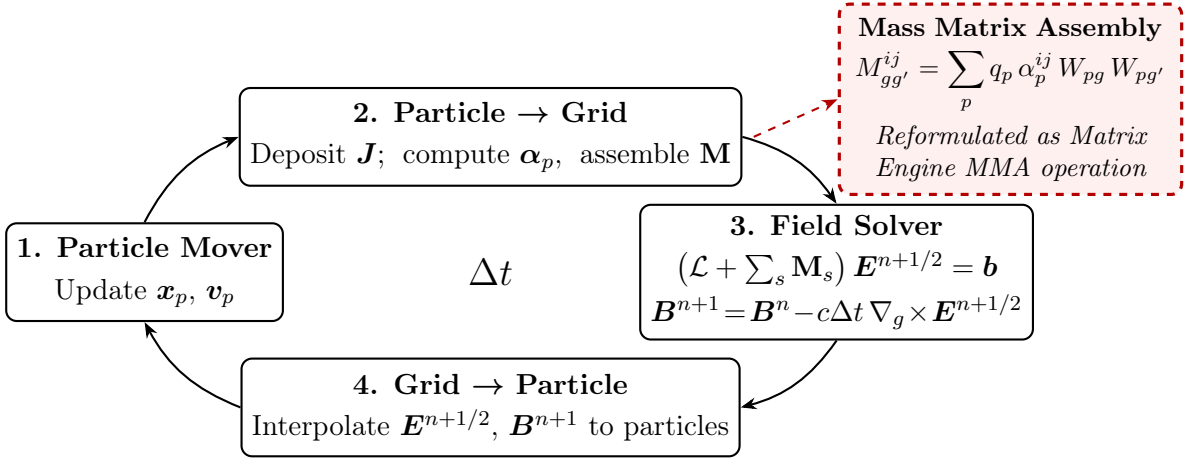
\begin{figure}[t!]
  \centering
\begin{tikzpicture}[
    box/.style={
        draw, rounded corners=4pt, minimum width=3.4cm, minimum height=1.3cm,
        align=center, font=\small, thick, fill=white
    },
    highlight/.style={
        draw=red!70!black, line width=1.2pt, dashed, rounded corners=4pt,
        inner sep=5pt, text width=4.2cm,
        align=center, font=\footnotesize, fill=red!6
    },
    arrow/.style={-{Stealth[length=6pt]}, thick},
    curvedarrow/.style={-{Stealth[length=6pt]}, thick},
]

\def\Rx{4.6cm}
\def\Ry{1.80cm}

\node[font=\large\bfseries] (center) at (0,0) {$\Delta t$};

\node[box] (mover) at ({-\Rx},0)
    {\textbf{1. Particle Mover}\\[2pt]
     Update $\bm{x}_p$, $\bm{v}_p$};

\node[box] (p2g) at (0,{\Ry})
    {\textbf{2. Particle $\boldsymbol{\to}$ Grid}\\[2pt]
     Deposit ${\bm{J}}$;\; compute
     $\bm{\alpha}_p$,\; assemble $\mathbf{M}$};

\node[box] (solver) at ({\Rx},0)
    {\textbf{3. Field Solver}\\[2pt]
     $\bigl(\mathcal{L}+\textstyle\sum_s\mathbf{M}_s\bigr)\,
      \bm{E}^{n+1/2}=\bm{b}$\\[2pt]
     $\bm{B}^{n+1}\!=\!\bm{B}^{n}\!-\!c\Delta t\,\nabla_g\!\times\!\bm{E}^{n+1/2}$};

\node[box] (g2p) at (0,{-\Ry})
    {\textbf{4. Grid $\boldsymbol{\to}$ Particle}\\[2pt]
     Interpolate $\bm{E}^{n+1/2}$, $\bm{B}^{n+1}$ to particles};

\draw[curvedarrow] (mover.north) to[bend left=20] (p2g.west);
\draw[curvedarrow] (p2g.east)    to[bend left=20] (solver.north);
\draw[curvedarrow] (solver.south) to[bend left=20] (g2p.east);
\draw[curvedarrow] (g2p.west)    to[bend left=20] (mover.south);

\node[highlight] (mm) at ({0.72*\Rx+3.6cm},{\Ry+0.5cm})
    {\textbf{Mass Matrix Assembly}\\[2pt]
     $M^{ij}_{gg'}=\displaystyle\sum_p q_p\,\alpha_p^{ij}\,W_{pg}\,W_{pg'}$\\[4pt]
     {\footnotesize\itshape Reformulated as Matrix Engine MMA operation}};

\draw[arrow, red!70!black, dashed] (p2g.east) ++(0.15,0) -- (mm.west);

\end{tikzpicture}
  \caption{Diagram of the PIC cycle for the ECSIM~\cite{lapenta2017ecsim} algorithm. The mass matrix calculation, the topic of this work, is highlighted in red.}
  \label{fig_pic_scheme}
\end{figure}

Figure~\ref{fig_pic_scheme} describes one cycle of the ECSIM algorithm. In the ECSIM method, the implicit coupling between the unknown electric field at time $n+1/2$, $\E^{n+1/2}$,  and the plasma current gives rise to a linear system for the field, defined on the grid nodes, of the form
\begin{equation}
  \left(\mathcal{L} + \sum_s \mathbf{M}_s\right) \E^{n+1/2} = \mathbf{b},
  \label{eq_field_eq}
\end{equation}
where $\mathcal{L}$ is a discrete curl-curl operator and $\mathbf{b}$ a right-hand-side vector depending on known quantities, such as current $\mathbf{J}$, at time $n$.
Once the electric field is known, the magnetic field is advanced via the discrete Faraday's law.
The operator $\mathbf{M}_s$ in Eq.~\ref{eq_field_eq} is the \emph{mass matrix} for species~$s$, its entries couple grid-node pairs $(g,g')$ through a $3\times 3$ tensor block:
\begin{equation}
  \bigl(\mathbf{M}_s\bigr)^{ij}_{gg'} = \frac{\beta_s}{c\, V_g} \sum_{p \in s} q_p\, \alpha_p^{ij}\, W_{pg}\, W_{pg'},
  \qquad i,j \in \{1,2,3\},
  \label{eq_mass_matrix_ecsim}
\end{equation}
where $q_p$ is the particle charge, $V_g$ is the cell volume, $\beta_s = q_s \Delta t / (2 m_s)$ is a species-dependent time-step parameter, and $\alpha_p^{ij}$ is the $(i,j)$ component of the particle rotation-response tensor $\bm\alpha_p$. The tensor $\bm\alpha_p$ is defined as
\begin{equation}
  \bm\alpha_p = \frac{1}{1 + \|\bm\omega_p\|^2}
  \left(\I - \mathcal{C}(\bm\omega_p) + \bm\omega_p \bm\omega_p^{\transpose} \right),
  \label{eq_alpha_matrix}
\end{equation}
where $\bm\omega_p = \beta_s\, \B(\x_p) / c$ is the dimensionless magnetization vector, with $\B(\x_p)$ being the magnetic field interpolated to the particle position, and $\mathcal{C}(\bm\omega)\,\bm u \equiv \bm\omega \times \bm u$ is the skew-symmetric cross-product operator.

Its assembly is the dominant cost of the implicit field solve, since for every particle one must evaluate and accumulate $N^2$ products of shape-function values, each multiplied by nine tensor components.

Dropping the species index for notational simplicity, the mass matrix can be written in the general form
\begin{equation}
  M^{ij}_{gg'} = \sigma\sum_{p=1}^{N_p} \tilde{s}_p^{ij}\, W_{pg}\, W_{pg'},
  \label{eq_mass_matrix_general}
\end{equation}
where $\sigma$ absorbs all constant prefactors and $\tilde{s}_p^{ij} \equiv q_p\, \alpha_p^{ij}$ is the per-particle coefficient tensor.
In the scalar case (e.g., the mass matrix in MPM~\cite{burgess1992_flip_mass_matrix}), one has $\tilde{s}_p^{ij} = q_p\, \delta^{ij}$ and only a single component per node pair.

\subsection{Mass matrix as a tensor contraction}
\label{sec:mm_matprod}

It has been recognized in the literature that the mass matrix computation, Eq.~\eqref{eq_mass_matrix_general}, admits a natural interpretation as a tensor contraction over the particle index~\cite{Montoya2022_fem_massMatrix,Perse2021_gempic}. Here we formalize this observation.

Consider a spatial grid with $N_g$ nodes and a system of $N_p$ particles.
The \emph{weight matrix} $\mat{W} \in \RR^{N_g \times N_p}$ collects all shape-function evaluations:
\begin{equation}
  W_{gp} \equiv W(\x_p - \x_g),
  \qquad g \in \{0,\ldots,N_g{-}1\},\; p \in \{0,\ldots,N_p{-}1\}.
  \label{eq_weight_matrix}
\end{equation}
For each tensor component $(i,j)$, we define the diagonal \emph{coefficient matrix}
\begin{equation}
  \mat{S}^{ij} = \mathrm{diag}\!\bigl(\tilde{s}_{0}^{ij},\, \ldots,\, \tilde{s}_{N_p-1}^{ij}\bigr) \in \RR^{N_p \times N_p}.
  \label{eq_S_diag}
\end{equation}
With Einstein summation, the mass matrix assembly in Eq.~\eqref{eq_mass_matrix_general} can be recast in a \emph{tensor contraction}
\begin{equation}
  M^{ij}_{gg'} = W_{gp}\, \tilde{s}_p^{ij}\, W_{g'p},
  \label{eq_D_entry}
\end{equation}
where the three factors are contracted over the repeated particle index~$p$.
In matrix notation, this reads
\begin{equation}
  \mat{M}^{ij} = \mat{W}\, \mat{S}^{ij}\, \mat{W}^{\transpose} \in \RR^{N_g \times N_g}.
  \label{eq_D_WSW}
\end{equation}
Although $\mat{M}^{ij} \in \RR^{N_g \times N_g}$ formally, it is extremely sparse since each particle contributes to at most $N \times N$ entries (with $N = (n{+}1)^d$), thus every row contains at most $(2n{+}1)^d$ nonzero entries. Additionally, the mass matrix is symmetric, for each fixed tensor component $(i,j)$,
\begin{equation}
  M^{ij}_{gg'} = M^{ij}_{g'g},
  \label{eq_spatial_symmetry}
\end{equation}
since the product $W_{gp}\, W_{g'p}$ in Eq.~\eqref{eq_D_entry} is invariant under interchange of~$g$ and~$g'$.
Each $N_g \times N_g$ block $\mat{M}^{ij}$ is therefore a real symmetric matrix. These symmetries, combined with the compact sparsity pattern, reduce the storage from $N_g^2$ entries to $\mathcal{O}(N_g)$ per component, indexed by canonical stencil offsets.

\subsection{Shape functions}
\label{sec:shape_functions}
The shape function $W(\x_p - \x_g)$ in Eq.~\eqref{eq_shape_function} assigns to each particle-node pair a non-negative interpolation weight that determines the coupling strength.
On a uniform Cartesian grid with spacing $\Delta x^\mu$ along coordinate $\mu \in \{1,\ldots,d\}$, the standard choice is a product of one-dimensional B-splines of order~$n$:
\begin{equation}
  W(\x_p - \x_g) = \prod_{\mu=1}^{d} \phi^{(n)}\!\left(\frac{x_p^\mu - x_g^\mu}{\Delta x^\mu}\right),
  \label{eq_shape_bspline}
\end{equation}
where $\phi^{(n)}$ is compactly supported on $[-(n{+}1)/2,\, (n{+}1)/2]$ and the support of each particle includes $N = (n{+}1)^d$ grid nodes.
The shape functions satisfy the partition-of-unity property $\sum_g W_{pg} = 1$ and are non-negative, ensuring conservative interpolation~\cite{birdsall1991_pic,Monaghan1985_particle_hydrodynamics}.

We consider the two cases of main practical relevance: first-order Cloud-In-Cell (CIC), with $N = 2^d$ support nodes per particle ($N=8$ in 3D), and second-order Triangular-Shaped Cloud (TSC), with $N = 3^d$ support nodes ($N=27$ in 3D).
For CIC, the stencil always coincides with the $2^d$ corner nodes of the cell containing the particle, so all particles in a cell share the same support.
For TSC, the identity of the support nodes depends on the particle position within the cell. In each dimension~$\mu$, the stencil is centered on the nearest grid node, so that for fractional coordinate $\xi_p^\mu \in [0,1)$ a particle with $\xi_p^\mu < 1/2$ uses support nodes $\{j{-}1, j, j{+}1\}$ (base offset $b_\mu = -1$), while $\xi_p^\mu \ge 1/2$ gives $\{j, j{+}1, j{+}2\}$ ($b_\mu = 0$).

\subsection{Tensor Core Architectures}
\label{sec:matrix_engines}

\begin{figure}[t!]
  \centering
  \begin{tikzpicture}[
    tileframe/.style={draw=black!70, line width=0.8pt},
    tilegrid/.style={draw=black!20, line width=0.35pt},
    dimm/.style={<->, line width=0.8pt, teal!70!black},
    dimn/.style={<->, line width=0.8pt, orange!85!black},
    dimk/.style={<->, line width=0.8pt, purple!65!black}
  ]
    \begin{scope}[shift={(0.0,0.0)}]
      \fill[gray!12] (0,0) rectangle (2.8,2.8);
      \draw[tileframe] (0,0) rectangle (2.8,2.8);
      \foreach \x in {0.7,1.4,2.1} {\draw[tilegrid] (\x,0) -- (\x,2.8);}
      \foreach \y in {0.7,1.4,2.1} {\draw[tilegrid] (0,\y) -- (2.8,\y);}
      \draw[dimm] (-0.30,0) -- (-0.30,2.8);
      \node[rotate=90, font=\scriptsize, text=teal!70!black] at (-0.58,1.4) {$M_t$};
      \draw[dimn] (0,2.98) -- (2.8,2.98);
      \node[font=\scriptsize, text=orange!85!black, anchor=south] at (1.4,3.05) {$N_t$};
    \end{scope}

    \node[font=\large] at (3.75,1.4) {$+$};

    \begin{scope}[shift={(4.80,0.0)}]
      \fill[teal!9] (0,0) rectangle (2.0,2.8);
      \draw[tileframe] (0,0) rectangle (2.0,2.8);
      \foreach \x in {0.5,1.0,1.5} {\draw[tilegrid] (\x,0) -- (\x,2.8);}
      \foreach \y in {0.7,1.4,2.1} {\draw[tilegrid] (0,\y) -- (2.0,\y);}
      \draw[dimm] (-0.30,0) -- (-0.30,2.8);
      \node[rotate=90, font=\scriptsize, text=teal!70!black] at (-0.58,1.4) {$M_t$};
      \draw[dimk] (0,2.98) -- (2.0,2.98);
      \node[font=\scriptsize, text=purple!65!black, anchor=south] at (1.0,3.05) {$K_t$};
    \end{scope}

    \node[font=\large] at (7.90,1.4) {$\times$};

    \begin{scope}[shift={(8.85,0.4)}]
      \fill[orange!10] (0,0) rectangle (2.8,2.0);
      \draw[tileframe] (0,0) rectangle (2.8,2.0);
      \foreach \x in {0.7,1.4,2.1} {\draw[tilegrid] (\x,0) -- (\x,2.0);}
      \foreach \y in {0.5,1.0,1.5} {\draw[tilegrid] (0,\y) -- (2.8,\y);}
      \draw[dimk] (-0.18,0) -- (-0.18,2.0);
      \node[rotate=90, font=\scriptsize, text=purple!65!black] at (-0.46,1.0) {$K_t$};
      \draw[dimn] (0,2.18) -- (2.8,2.18);
      \node[font=\scriptsize, text=orange!85!black, anchor=south] at (1.4,2.25) {$N_t$};
    \end{scope}
  \end{tikzpicture}
  \caption{Abstract matrix-engine MMA update written in tile form as $\mat{D} \leftarrow \mat{D} + \mat{A}\mat{B}$. The accumulator has size $M_t \times N_t$, the left operand has size $M_t \times K_t$, and the right operand has size $K_t \times N_t$.}
  \label{fig_mma_schematic}
\end{figure}
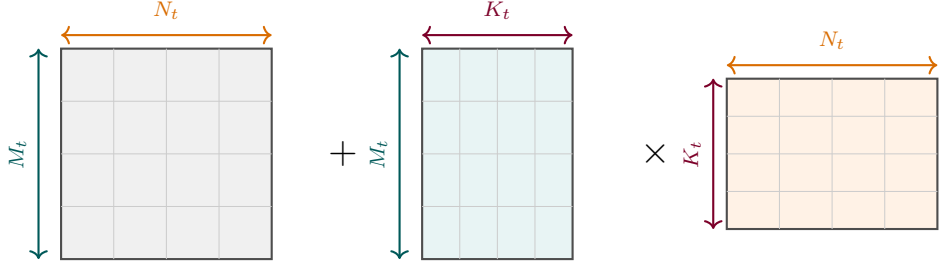

For the scope of this work, we abstract a \emph{tensor core} as any hardware unit that realizes the following operations.
Given fixed positive integers $M_t$, $N_t$, $K_t$, which define the \emph{tile shape}, the engine accepts operand tiles $A \in \mathcal{F}_{\mathrm{in}}^{M_t \times K_t}$ and $B \in \mathcal{F}_{\mathrm{in}}^{K_t \times N_t}$, where $\mathcal{F}_{\mathrm{in}}$ is the floating-point format of the input operands, and updates an accumulator tile $D \in \mathcal{F}_{\mathrm{acc}}^{M_t \times N_t}$, stored in a (generally wider) accumulation format $\mathcal{F}_{\mathrm{acc}} \supseteq \mathcal{F}_{\mathrm{in}}$, according to the MMA rule
\begin{equation}
  D_{ab} \leftarrow D_{ab} + A_{ak}\, B_{kb},
  \qquad a \in \{0,\ldots,M_t{-}1\},\; b \in \{0,\ldots,N_t{-}1\},
  \label{eq_mma_op}
\end{equation}
where summation over the repeated index $k \in \{0,\ldots,K_t{-}1\}$ is implied. Figure~\ref{fig_mma_schematic} summarizes this abstraction in the compact form of the MMA update $\mat{D} \leftarrow \mat{D} + \mat{A}\mat{B}$.
Each invocation of Eq.~\eqref{eq_mma_op} represents a tiled matrix multiply-accumulate involving $2\, M_t\, N_t\, K_t$ floating-point operations.This operation is exposed as a single MMA instruction in the programming model and is executed on specialized tensor-core hardware, yielding higher throughput than an equivalent implementation built from scalar FMA instructions~\cite{markidis2019_tensorcores}. 

Additionally, the use of a reduced-precision input format $\mathcal{F}_{\mathrm{in}}$ (such as FP16, TF32, or BF16) with a wider accumulation format $\mathcal{F}_{\mathrm{acc}}$ (such as FP32 or FP64) allows the hardware to maximize throughput for the multiply stage while preserving numerical accuracy in the accumulation.

All major contemporary computer architectures provide such tensor cores: NVIDIA \emph{tensor cores}~\cite{markidis2019_tensorcores}, AMD \emph{matrix cores}~\cite{schieffer2024_matrixcores}, Intel \emph{Advanced Matrix Extensions} (AMX)~\cite{kim2024_intel_amx}, and Google \emph{Tensor Processing Units} (TPUs)~\cite{jouppi2017_tpu}.
These implementations differ in their supported tile shapes $(M_t, N_t, K_t)$, and precision formats $(\mathcal{F}_{\mathrm{in}}, \mathcal{F}_{\mathrm{acc}})$, but all conform to the abstract MMA interface defined by Eq.~\eqref{eq_mma_op}.

\section{Methodology}
\label{sec:methodology}

In Section~\ref{sec:mm_matprod} we recall that the mass matrix assembly is a tensor contraction $M^{ij}_{gg'} = W_{gp}\, \tilde{s}_p^{ij}\, W_{g'p}$ over the particle index~$p$ (Eq.~\eqref{eq_D_entry}), and that the compact support of the shape functions restricts each particle's contribution to a small block of~$\mat{M}^{ij}$.
In this section, we show how we leverage the inherent sparsity of~$\mat{M}^{ij}$ to decompose its calculation at the cell level and map the tensor contraction onto the fixed-size tile operations provided by hardware matrix engines.
The derivation is carried out in full generality, independent of the interpolation order, the number of spatial dimensions, the scalar or tensorial nature of the per-particle coefficient, and the tile shapes of the matrix engine.

\subsection{Cell-local tensor contraction}
\label{sec:cell_local}
Firstly, wo factor the three-tensor product in Eq.~\eqref{eq_D_entry} into a two-operand matrix multiply by absorbing the diagonal coefficient into the left factor:
\begin{equation}
  M^{ij}_{gg'} = A^{ij}_{gp}\, B_{pg'}, \qquad \mat{A}^{ij} = \mat{W}\, \mat{S}^{ij} \in \RR^{N_g \times N_p}, \;\; \mat{B} = \mat{W}^{\transpose} \in \RR^{N_p \times N_g}.
  \label{eq_D_AB}
\end{equation}

\begin{figure}[t!]
  \centering
    \begin{tikzpicture}[
    >=Latex,
    scale=0.92,
    transform shape,
    mtx/.style={
      matrix of math nodes,
      left delimiter={[},
      right delimiter={]},
      nodes={
        font=\scriptsize,
        anchor=center,
        inner sep=1.6pt,
        text depth=0.25ex,
        text height=1.6ex
      },
      row sep=4pt,
      column sep=4pt
    },
    pzero/.style={draw=teal!70!black, rounded corners=2pt, line width=0.85pt},
    pone/.style={draw=orange!85!black, rounded corners=2pt, line width=0.85pt},
    ptwo/.style={draw=red!75!black, rounded corners=2pt, line width=0.85pt},
    plast/.style={draw=black!70, dashed, rounded corners=2pt, line width=0.85pt}
  ]
    \matrix (A) [mtx] {
      W_{g_0p_0}\tilde{s}_{p_0}^{ij} & W_{g_0p_1}\tilde{s}_{p_1}^{ij} & \cdots & W_{g_0p_N}\tilde{s}_{p_N}^{ij} \\
      W_{g_1p_0}\tilde{s}_{p_0}^{ij} & W_{g_1p_1}\tilde{s}_{p_1}^{ij} & \cdots & W_{g_1p_N}\tilde{s}_{p_N}^{ij} \\
      \vdots & \vdots & \ddots & \vdots \\
      W_{g_{N_g-1}p_0}\tilde{s}_{p_0}^{ij} & W_{g_{N_g-1}p_1}\tilde{s}_{p_1}^{ij} & \cdots & W_{g_{N_g-1}p_N}\tilde{s}_{p_N}^{ij} \\
    };

    \node[pzero, fit=(A-1-1)(A-4-1), inner sep=2pt] (Acol0) {};
    \node[pone,  fit=(A-1-2)(A-4-2), inner sep=2pt] (Acol1) {};
    \node[plast, fit=(A-1-4)(A-4-4), inner sep=2pt] (AcolN) {};

    \node[font=\scriptsize, text=teal!70!black, above=2pt of Acol0] {$p_0$};
    \node[font=\scriptsize, text=orange!85!black, above=2pt of Acol1] {$p_1$};
    \node[font=\scriptsize, above=2pt of AcolN] {$p_N$};

    \node[right=8mm of A] (times) {$\times$};

    \matrix (B) [mtx, right=8mm of times] {
      W_{g_0p_0} & W_{g_1p_0} & \cdots & W_{g_{N_g-1}p_0} \\
      W_{g_0p_1} & W_{g_1p_1} & \cdots & W_{g_{N_g-1}p_1} \\
      \vdots & \vdots & \ddots & \vdots \\
      W_{g_0p_N} & W_{g_1p_N} & \cdots & W_{g_{N_g-1}p_N} \\
    };

    \node[pzero, fit=(B-1-1)(B-1-4), inner sep=2pt] (Brow0) {};
    \node[pone,  fit=(B-2-1)(B-2-4), inner sep=2pt] (Brow1) {};
    \node[plast, fit=(B-4-1)(B-4-4), inner sep=2pt] (BrowN) {};

    \node[font=\scriptsize, text=teal!70!black, anchor=east] at ($(Brow0.west)+(-8pt,0)$) {$p_0$};
    \node[font=\scriptsize, text=orange!85!black, anchor=east] at ($(Brow1.west)+(-8pt,0)$) {$p_1$};
    \node[font=\scriptsize, anchor=east] at ($(BrowN.west)+(-8pt,0)$) {$p_N$};
\end{tikzpicture}
  \caption{Schematic of the factorization $\mat{M}^{ij} = \mat{A}^{ij}\mat{B}$. Each column of $\mat{A}^{ij}$ corresponds to one particle and contains that particle's interpolation weights scaled by $\tilde{s}_p^{ij}$. The matching row of $\mat{B}$ contains the same particle weights unscaled. The matrix product contracts over the shared particle index and sums the per-particle outer products into the mass matrix.}
  \label{fig_AB_factorization}
\end{figure}
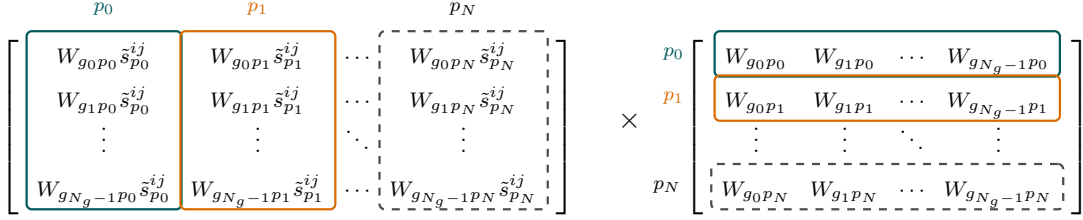

Figure~\ref{fig_AB_factorization} visualizes this factorization: column~$p$ of $\mat{A}^{ij}$ stores the interpolation weights of particle~$p$ scaled by its coefficient $\tilde{s}_p^{ij}$, while row~$p$ of $\mat{B}$ stores the same particle weights without the scaling. The product $\mat{A}^{ij}\mat{B}$ therefore contracts over the shared particle index and accumulates the outer-product contribution of each particle into the grid-grid matrix.\\

Given the compact support of the shape function $W(\x_p - \x_g)$, Eq. \ref{eq_shape_bspline}, each column (row) $p$ in the global matrix $\mat{A}^{ij}$ ($\mat{B}$) would have at most $(n+1)^d$ non zero entries, leading to an highly sparse matrix-matrix multiplication. We can therefore leverage the regularity and compactness of the shape functions support to decompose the mass matrix assembly in a series of \emph{cell-local} tensor contractions rather than a single large global operation.

We assume that particles have been sorted by cell, which is typically the case for production-level simulations~\cite{bowers2001accelerating}.
Let a cell contain $P$ particles and let $\mathcal{N} = \{n_0, \ldots, n_{N-1}\}$ be a fixed set of $N$ grid nodes such that the support of every particle in the cell is contained in~$\mathcal{N}$.
Restricting the global weight tensor (Eq.~\eqref{eq_weight_matrix}) and coefficient tensor (Eq.~\eqref{eq_S_diag}) to these $P$ particles and $N$ nodes yields cell-local matrices $\hat{\mat{W}} \in \RR^{N \times P}$ and $\hat{\mat{S}}^{ij} \in \RR^{P \times P}$.
The global tensor contraction (Eq.~\eqref{eq_D_entry}) then applies block-wise and the cell-local mass matrix is
\begin{equation}
  \hat{M}^{ij}_{ab} = \hat{W}_{ap}\, \hat{s}_p^{ij}\, \hat{W}_{bp}.
  \label{eq_cell_local_einstein}
\end{equation}

\subsection{Particle batching for MMA tiles}
\label{sec:batch_decomp}

The cell-local tensor contraction in Eq.~\eqref{eq_cell_local_einstein} sums over all $P$ particles in a cell, but the hardware MMA tile contracts over a fixed inner dimension~$K_t$. Thus, we partition the $P$ particles into $\lceil P/K_t \rceil$ consecutive \emph{batches} of $K_t$ particles each, with the last batch zero-padded if $K_t \nmid P$.

For batch $\beta$ containing particles $\{p_{\beta,0}, \ldots, p_{\beta,K_t-1}\}$, we define the batch weight matrix
\begin{equation}
  \W^{(\beta)}_{ak} \equiv \hat{W}_{a,p_{\beta,k}}, \quad
  \boldsymbol{\W^{(\beta)}} \in \RR^{N \times K_t},
  \label{eq_Wbeta}
\end{equation}
and the corresponding diagonal coefficient slice
\begin{equation}
  \boldsymbol{\mat{S}^{ij,(\beta)}} = \mathrm{diag}\!\bigl(\hat{s}^{ij}_{p_{\beta,0}},\, \ldots,\, \hat{s}^{ij}_{p_{\beta,K_t-1}}\bigr) \in \RR^{K_t \times K_t}.
  \label{eq_Sbeta}
\end{equation}
The MMA operands for batch $\beta$ are
\begin{equation}
  A^{ij,(\beta)}_{ak} \equiv \W^{(\beta)}_{ak}\, \hat{s}^{ij}_{p_{\beta,k}},
  \qquad
  B^{(\beta)}_{kb} \equiv \W^{(\beta)}_{bk},
  \label{eq_AB_batch}
\end{equation}
with $\mat{A}^{ij,(\beta)} \in \RR^{N \times K_t}$ and $\mat{B}^{(\beta)} \in \RR^{K_t \times N}$.
The full cell-local mass matrix is the sum of per-batch products
\begin{equation}
  \hat{M}^{ij}_{ab} = \sum_{\beta=0}^{\lceil P/K_t \rceil - 1}
  A^{ij,(\beta)}_{ak}\, B^{(\beta)}_{kb},
  \label{eq_D_batches}
\end{equation}
crucially, this summation maps exactly onto the hardware MMA instruction defined in Eq.~\eqref{eq_mma_op}.
Initializing the accumulator tile to $D^{ij}_{ab} \leftarrow 0$, each batch $\beta$ triggers the in-place update
\begin{equation}
  D^{ij}_{ab} \leftarrow D^{ij}_{ab} + A^{ij,(\beta)}_{ak}\, B^{(\beta)}_{kb},
  \label{eq_mma_accumulate}
\end{equation}
thus, after all $\lceil P/K_t \rceil$ MMA calls the accumulator holds the exact cell-local mass matrix: $D^{ij}_{ab} = \hat{M}^{ij}_{ab}$.
The mathematical sum over particle batches in Eq.~\eqref{eq_D_batches} is therefore realized by a loop of hardware MMA instructions that accumulate in place, requiring no intermediate storage and no explicit reduction step.

When $N > M_t$ (or $N > N_t$), the $N \times N$ accumulator is covered by $\lceil N/M_t \rceil \times \lceil N/N_t \rceil$ tiles, each executing an independent MMA instruction per batch.
The weight matrix rows are then padded to the nearest multiple of $M_t$ to fill incomplete tiles.

\subsection{Support-group decomposition}
\label{sec:support_groups}

The cell-local contraction in Eq.~\eqref{eq_cell_local_einstein} assumes that all particles
contributing in a given product share the same support nodes. For shape functions of order $n \ge 2$, the stencil placement depends on the particle position within the cell, so two particles in the same cell may touch different (though overlapping) subsets of $(n{+}1)^d$ grid nodes.

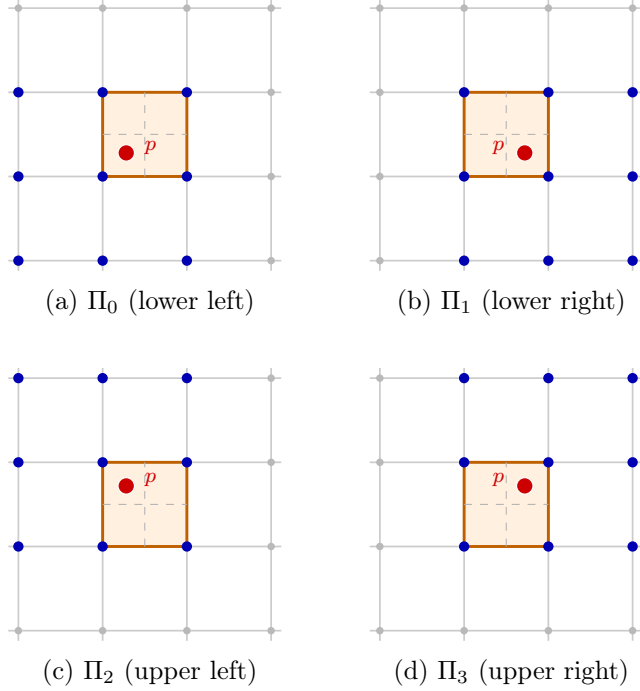
\begin{figure}[t!]
  \centering
  \subfloat[$\Pi_0$ (lower left)]{
    \begin{tikzpicture}[
      x=1.12cm,
      y=1.12cm,
      gridline/.style={line width=0.65pt, gray!40},
      gridnode/.style={circle, fill=gray!55, inner sep=1.0pt},
      particle/.style={circle, fill=red!80!black, draw=white, line width=0.45pt, inner sep=2.2pt}
    ]
      \path[use as bounding box] (-1.18,-1.18) rectangle (2.18,2.18);
      \fill[orange!12] (0,0) rectangle (1,1);
      \foreach \x in {-1,0,1,2} {
        \draw[gridline] (\x,-1.12) -- (\x,2.12);
      }
      \foreach \y in {-1,0,1,2} {
        \draw[gridline] (-1.12,\y) -- (2.12,\y);
      }
      \draw[line width=1.1pt, orange!75!black] (0,0) rectangle (1,1);
      \draw[dashed, gray!60] (0.5,0) -- (0.5,1);
      \draw[dashed, gray!60] (0,0.5) -- (1,0.5);
      \foreach \ix in {-1,0,1,2} {
        \foreach \iy in {-1,0,1,2} {
          \node[gridnode] at (\ix,\iy) {};
        }
      }
      \foreach \ix in {-1,0,1} {
        \foreach \iy in {-1,0,1} {
          \fill[blue!70!black] (\ix,\iy) circle (1.95pt);
        }
      }
      \node[particle] at (0.28,0.28) {};
      \node[font=\scriptsize, red!80!black, anchor=west] at (0.37,0.35) {$p$};
\end{tikzpicture}  
  }
  \hspace{14pt}
  \subfloat[$\Pi_1$ (lower right)]{
    \begin{tikzpicture}[
      x=1.12cm,
      y=1.12cm,
      gridline/.style={line width=0.65pt, gray!40},
      gridnode/.style={circle, fill=gray!55, inner sep=1.0pt},
      particle/.style={circle, fill=red!80!black, draw=white, line width=0.45pt, inner sep=2.2pt}
    ]
      \path[use as bounding box] (-1.18,-1.18) rectangle (2.18,2.18);
      \fill[orange!12] (0,0) rectangle (1,1);
      \foreach \x in {-1,0,1,2} {
        \draw[gridline] (\x,-1.12) -- (\x,2.12);
      }
      \foreach \y in {-1,0,1,2} {
        \draw[gridline] (-1.12,\y) -- (2.12,\y);
      }
      \draw[line width=1.1pt, orange!75!black] (0,0) rectangle (1,1);
      \draw[dashed, gray!60] (0.5,0) -- (0.5,1);
      \draw[dashed, gray!60] (0,0.5) -- (1,0.5);
      \foreach \ix in {-1,0,1,2} {
        \foreach \iy in {-1,0,1,2} {
          \node[gridnode] at (\ix,\iy) {};
        }
      }
      \foreach \ix in {0,1,2} {
        \foreach \iy in {-1,0,1} {
          \fill[blue!70!black] (\ix,\iy) circle (1.95pt);
        }
      }
      \node[particle] at (0.72,0.28) {};
      \node[font=\scriptsize, red!80!black, anchor=west] at (0.21,0.35) {$p$};
\end{tikzpicture}
  }
  
  \vspace{6pt}
  \subfloat[$\Pi_2$ (upper left)]{
    \begin{tikzpicture}[
      x=1.12cm,
      y=1.12cm,
      gridline/.style={line width=0.65pt, gray!40},
      gridnode/.style={circle, fill=gray!55, inner sep=1.0pt},
      particle/.style={circle, fill=red!80!black, draw=white, line width=0.45pt, inner sep=2.2pt}
    ]
      \path[use as bounding box] (-1.18,-1.18) rectangle (2.18,2.18);
      \fill[orange!12] (0,0) rectangle (1,1);
      \foreach \x in {-1,0,1,2} {
        \draw[gridline] (\x,-1.12) -- (\x,2.12);
      }
      \foreach \y in {-1,0,1,2} {
        \draw[gridline] (-1.12,\y) -- (2.12,\y);
      }
      \draw[line width=1.1pt, orange!75!black] (0,0) rectangle (1,1);
      \draw[dashed, gray!60] (0.5,0) -- (0.5,1);
      \draw[dashed, gray!60] (0,0.5) -- (1,0.5);
      \foreach \ix in {-1,0,1,2} {
        \foreach \iy in {-1,0,1,2} {
          \node[gridnode] at (\ix,\iy) {};
        }
      }
      \foreach \ix in {-1,0,1} {
        \foreach \iy in {0,1,2} {
          \fill[blue!70!black] (\ix,\iy) circle (1.95pt);
        }
      }
      \node[particle] at (0.28,0.72) {};
      \node[font=\scriptsize, red!80!black, anchor=west] at (0.37,0.79) {$p$};
\end{tikzpicture}
  }
  \hspace{14pt}
  \subfloat[$\Pi_3$ (upper right)]{
    \begin{tikzpicture}[
      x=1.12cm,
      y=1.12cm,
      gridline/.style={line width=0.65pt, gray!40},
      gridnode/.style={circle, fill=gray!55, inner sep=1.0pt},
      particle/.style={circle, fill=red!80!black, draw=white, line width=0.45pt, inner sep=2.2pt}
    ]
      \path[use as bounding box] (-1.18,-1.18) rectangle (2.18,2.18);
      \fill[orange!12] (0,0) rectangle (1,1);
      \foreach \x in {-1,0,1,2} {
        \draw[gridline] (\x,-1.12) -- (\x,2.12);
      }
      \foreach \y in {-1,0,1,2} {
        \draw[gridline] (-1.12,\y) -- (2.12,\y);
      }
      \draw[line width=1.1pt, orange!75!black] (0,0) rectangle (1,1);
      \draw[dashed, gray!60] (0.5,0) -- (0.5,1);
      \draw[dashed, gray!60] (0,0.5) -- (1,0.5);
      \foreach \ix in {-1,0,1,2} {
        \foreach \iy in {-1,0,1,2} {
          \node[gridnode] at (\ix,\iy) {};
        }
      }
      \foreach \ix in {0,1,2} {
        \foreach \iy in {0,1,2} {
          \fill[blue!70!black] (\ix,\iy) circle (1.95pt);
        }
      }
      \node[particle] at (0.72,0.72) {};
      \node[font=\scriptsize, red!80!black, anchor=west] at (0.21,0.79) {$p$};
\end{tikzpicture}
  }
  \caption{Four possible two-dimensional TSC supports inside a fixed cell. The orange cell is the particle-containing cell, the dashed lines mark its half-cell boundaries, and the blue dots are the $3 \times 3$ interpolation nodes used for the particle position shown in each panel. The four panels correspond to the lower-left, lower-right, upper-left, and upper-right particle classes within the cell. Particles that fall in the same panel share the same node set and therefore belong to the same group.}
  \label{fig_support_group_decomposition}
\end{figure}

Let $\mathcal P_c = \{p_0,\ldots,p_{P-1}\}$ be the particles in cell $c$.
We partition $\mathcal P_c$ into $G$ groups $\Pi_0,\ldots,\Pi_{G-1}$ by grouping particles that share identical support nodes:
\begin{equation}
  \mathcal P_c = \bigsqcup_{\gamma=0}^{G-1} \Pi_\gamma,
  \qquad
  \forall\, p \in \Pi_\gamma:\; \mathcal N_p = \mathcal N_\gamma ,
  \label{eq_group_partition}
\end{equation}
where $\mathcal N_\gamma$ denotes the common support of group $\Pi_\gamma$ and $|\Pi_\gamma| = P_\gamma$.
Figure~\ref{fig_support_group_decomposition} illustrates this decomposition for second-order interpolation in two dimensions, showing the four possible $3 \times 3$ TSC nodal supports inside one cell.\\

The matrix product Eq.~\eqref{eq_cell_local_einstein} applies independently within each group.
We define the per-group weight matrix $\mat{W}^{(\gamma)} \in \RR^{N \times P_\gamma}$ restricted to the particles in $\Pi_\gamma$ and the nodes in $\mathcal{N}_\gamma$.
The per-group mass matrix is
\begin{equation}
  M^{ij,(\gamma)}_{ab} = W^{(\gamma)}_{ap}\, \tilde{s}_p^{ij}\, W^{(\gamma)}_{bp},
  \label{eq_M_group}
\end{equation}
and the full cell contribution to the global mass matrix is the sum over all groups:
\begin{equation}
 \hat{M}^{ij}_{ab} \;\mathrel{+}=\; \sum_{\gamma=0}^{G-1} M^{ij,(\gamma)}_{ab},
  \label{eq_M_all_groups}
\end{equation}
where each $\mat{M}^{ij,(\gamma)}$ is deposited to the (generally distinct) global nodes in $\mathcal{N}_\gamma$.

An alternative is to embed all particles' weights into a single vector of length $|\bigcup_\gamma \mathcal{N}_\gamma|$, padding with zeros for unsupported nodes, and form one large outer product.
This is mathematically correct since zero weights remove cross-terms, but computationally wasteful, as the accumulator grows from $N^2$ to $|\bigcup_\gamma \mathcal{N}_\gamma|^2$ with many structurally zero entries.

The number of support groups $G$ depends on the interpolation order $n$:
\begin{itemize}
  \item First-order (CIC): $G = 1$. All particles in a cell share the same $(n{+}1)^d = 2^d$ nodes.
  \item Second-order (TSC): $G \le n^d = 2^d$ in $d$ dimensions. Each particle's stencil can be shifted by one node per dimension relative to the cell corner.
  \item In general, for order~$n$ in $d$ dimensions: $G \le n^d$.
\end{itemize}

The complete mass matrix assembly at the cell level thus has a two-level structure:
\begin{enumerate}
  \item \emph{Support groups} ($\gamma = 0,\ldots,G{-}1$): partition particles by the set of $N$ grid nodes they touch.
  \item \emph{Batches of $K_t$}: within each support group, particles are further partitioned into batches of size $K_t$ for the MMA tile operation.
\end{enumerate}
After each support group is processed, the accumulated tile(s) are deposited to the global mass matrix at the addresses determined by the group's node set $\mathcal{N}_\gamma$.

\subsection{Mass matrix sparse stencil deposition}
\label{sec:stencil}

Once the cell-local (or per-group) mass matrix $\hat{M}^{ij}_{ab}$ has been accumulated, it must be scattered into the global mass matrix, which is stored in a compact sparse format indexed by canonical stencil offsets.

For an order-$n$ shape function in $d$ dimensions, the displacement between any two nodes in a particle's support ranges over $\{-n, \ldots, +n\}^d$, giving $(2n{+}1)^d$ possible offsets.
By exploiting the symmetry $M^{ij}_{gg'} = M^{ji}_{g'g}$, which reduces to $M_{gg'} = M_{g'g}$ in the scalar case, only the "forward half" plus the diagonal need be stored.
Concretely, in the case of first and second order interpolation functions, we have:
\begin{itemize}
  \item CIC ($n=1$): displacements in $\{-1,0,+1\}^d$, giving $3^d$ offsets, of which $(3^d{+}1)/2$ are canonical.
  \item TSC ($n=2$): displacements in $\{-2,\ldots,+2\}^d$, giving $5^d$ offsets, of which $(5^d{+}1)/2$ are canonical.
\end{itemize}

When support groups are present ($G > 1$), different groups deposit to different, but overlapping, sets of global nodes.
The canonical stencil index for a given local entry $(a,b)$ therefore depends on the group's base offset.
A precomputed lookup table can be used to map each local node pair to the corresponding canonical stencil index and global node address, ensuring correct assembly irrespective of the number of groups.

\subsection{Algorithmic summary}
\label{sec:algorithm}

Algorithm~\ref{alg:general} describes the general cell-level procedure for assembling the mass matrix using hardware matrix engines with generic MMA tiles of shape $(M_t, N_t, K_t)$, with $G$ support groups per cell and $N_c$ tensor components per node pair.
For first-order shape functions ($n=1$), only a single support group exists ($G=1$) and the outer loop is trivial.
For higher-order shape functions ($n \ge 2$), the support-group loop executes up to $G = n^d$ iterations.
Within each group, the batch loop processes all assigned particles in chunks of $K_t$.

\begin{algorithm}[h!]
\caption{Cell-local mass matrix assembly via tensor cores.}
\label{alg:general}
\begin{algorithmic}[1]
\State \textbf{Input:} Cell particle list $\{p_0,\ldots,p_{P-1}\}$, grid geometry, interpolation order $n$, tile shape $(M_t, N_t, K_t)$.
\State \textbf{Output:} Contributions accumulated into global mass matrix $\mat{M}^{ij}$.
\Statex
\State $N \leftarrow (n{+}1)^d$ \Comment{Nodes per particle support}
\State $N_{\mathrm{pad}} \leftarrow \lceil N / M_t \rceil \cdot M_t$ \Comment{Padded node count}
\State $T_r \leftarrow N_{\mathrm{pad}} / M_t$,\; $T_c \leftarrow N_{\mathrm{pad}} / N_t$ \Comment{Tile grid}
\Statex
\For{each support group $\gamma \in \{0,\ldots,G{-}1\}$}
  \State Determine node set $\mathcal{N}_\gamma$ and deposit addresses.
  \State Initialize accumulator tiles: $D^{ij}_{rc} \leftarrow 0$,\; $r \in \{0,\ldots,T_r{-}1\}$,\; $c \in \{0,\ldots,T_c{-}1\}$.
  \Statex
  \For{each particle $p \in \Pi_\gamma$}
    \State Compute shape-function weights $W_{ap}$,\; $a \in \{0,\ldots,N{-}1\}$.
    \State Compute per-particle coefficients $\tilde{s}_p^{ij}$.
    \State Buffer $W_{ap}$ and $\tilde{s}_p^{ij}$ into current batch.
    \If{batch full ($K_t$ particles accumulated)}
      \State Form $A^{ij,(\beta)}_{ak} = W_{a,p_{\beta,k}}\, \tilde{s}^{ij}_{p_{\beta,k}}$ and $B^{(\beta)}_{kb} = W_{b,p_{\beta,k}}$.
      \For{each tile $(r,c)$ and each component $(i,j)$}
        \State \textbf{MMA:}\; $D^{ij}_{rc} \leftarrow D^{ij}_{rc} + A^{ij,(\beta)}_{r}\, B^{(\beta)}_{c}$
      \EndFor
    \EndIf
  \EndFor
  \State Process remaining partial batch with zero-padded MMA.
  \State Deposit $D^{ij}_{rc}$ to global $\mat{M}^{ij}$ via stencil lookup.
\EndFor
\end{algorithmic}
\end{algorithm}

\pagebreak

\section{Numerical results}
\label{sec:results}
To demonstrate the practical viability of the mass matrix matrix-product reformulation, we specialize the general framework of Section~\ref{sec:methodology} to first-order (CIC) and second-order (TSC) B-spline interpolation in three dimensions and implement it on NVIDIA GPUs using the Warp Matrix Multiply-Accumulate (WMMA) intrinsics provided by the CUDA programming model.
We target two tile formats:

\begin{itemize}
  \item \textbf{FP64 tile $(M_t, N_t, K_t) = (8, 8, 4)$}: all operands in double precision.
    The $8 \times 8$ accumulator matches the CIC support size exactly ($N = 8$), so a single tile covers the full outer product with batch size $K_t = 4$.
  \item \textbf{TF32 tile $(M_t, N_t, K_t) = (16, 16, 8)$}: inputs in TF32 (10-bit mantissa, 8-bit exponent) with FP32 accumulation.
    For TSC ($N = 27$), the weight matrix is padded to $32$ rows and the $32 \times 32$ accumulator is covered by $2 \times 2 = 4$ tiles, with batch size $K_t = 8$.
    Exploiting the spatial symmetry $\hat{M}_{ab} = \hat{M}_{ba}$ (Eq.~\eqref{eq_spatial_symmetry}), tile $(1,0)$ is skipped and only the three upper-triangle tiles are computed.
\end{itemize}

Both kernels assign one warp per cell with a grid-stride loop over cells.
Particles are processed in bounded chunks (up to $P_{\max} = 64$) to control register pressure, and accumulator fragments persist across chunks to reduce atomic reductions in main memory.
For CIC, $8\times K_t$-sized batches are assembled via warp shuffles. For TSC, fragment data is staged through shared memory to assemble the $16 \times K_t$ sub-tiles.
Table~\ref{tab:mma_summary} summarizes the mapping from interpolation order to WMMA tile parameters in the experiments.

Firstly, we assess the performance of MMA mass matrix assembly in isolation, comparing the execution times of the WMMA kernels with those of conventional, highly optimized GPU implementations. Then, we assess the benefit provided by MMA in a production PIC simulation, using the ECSIM algorithm. We run all the isolation experiments on a single node machine, equipped with an AMD EPYC 7302P 32-core CPU, and an NVIDIA A100 GPU. The production PIC simulations are run on a multi-node cluster equipped with 2x AMD Rome 7H12 CPUs and 4x NVIDIA A100 GPUs per node.

\begin{table}[h!]
\centering
\caption{Mapping of cell-local mass matrix dimensions to MMA tile parameters for the NVIDIA tensor core implementation. $N$ is the number of nodes in a particle's support, $M_t$, $N_t$, and $K_t$ are the tile size, with $K_t$ corresponding to the batch size, and $G$ is the number of support groups per cell. $Padded$ is the matrix size after rounding up, while $Tiles$ represents the number of tiles required to cover the matrix of size $Padded$.}
\label{tab:mma_summary}
\begin{tabular}{lcccccc}
\toprule
Interpolation & Precision & $N$ & Padded & Tile $(M_t, N_t, K_t)$ & Tiles & $G$ \\
\midrule
CIC ($n=1$) & FP64 & 8  & 8  & $(8, 8, 4)$  & 1 & 1  \\
TSC ($n=2$) & TF32/FP32 & 27 & 32 & $(16, 16, 8)$ & $3^\dagger$ & $\le 8$ \\
\bottomrule
\multicolumn{6}{l}{\footnotesize $^\dagger$\,Upper-triangle tiles only: (0,0), (0,1), (1,1). Tile (1,0) is skipped by symmetry.}
\end{tabular}
\end{table}

\subsection{WMMA comparison against optimized conventional GPU kernels}
\label{sec:comparison}
The isolation experiments measure the mass matrix assembly kernel on a 3D domain of $N_x \times N_y \times N_z$ cells with a single species and a uniform number of particles per cell (ppc), pre-sorted by cell.
We test both CIC and TSC interpolation for both scalar and $3\times 3$ ECSIM tensorial mass matrices (Eq.~\ref{eq_mass_matrix_ecsim}). CIC experiments use FP64 data format, while TSC experiments use FP32 data with TF32 inputs and FP32 accumulation.
Two parameter investigations are performed: (i) varying ppc with a fixed $16 \times 16 \times 16$ grid, and (ii) varying the grid size with a fixed 128~ppc. 

\begin{figure}[t!]
    \centering
    \subfloat[]{
    \includegraphics[width=0.48\linewidth]{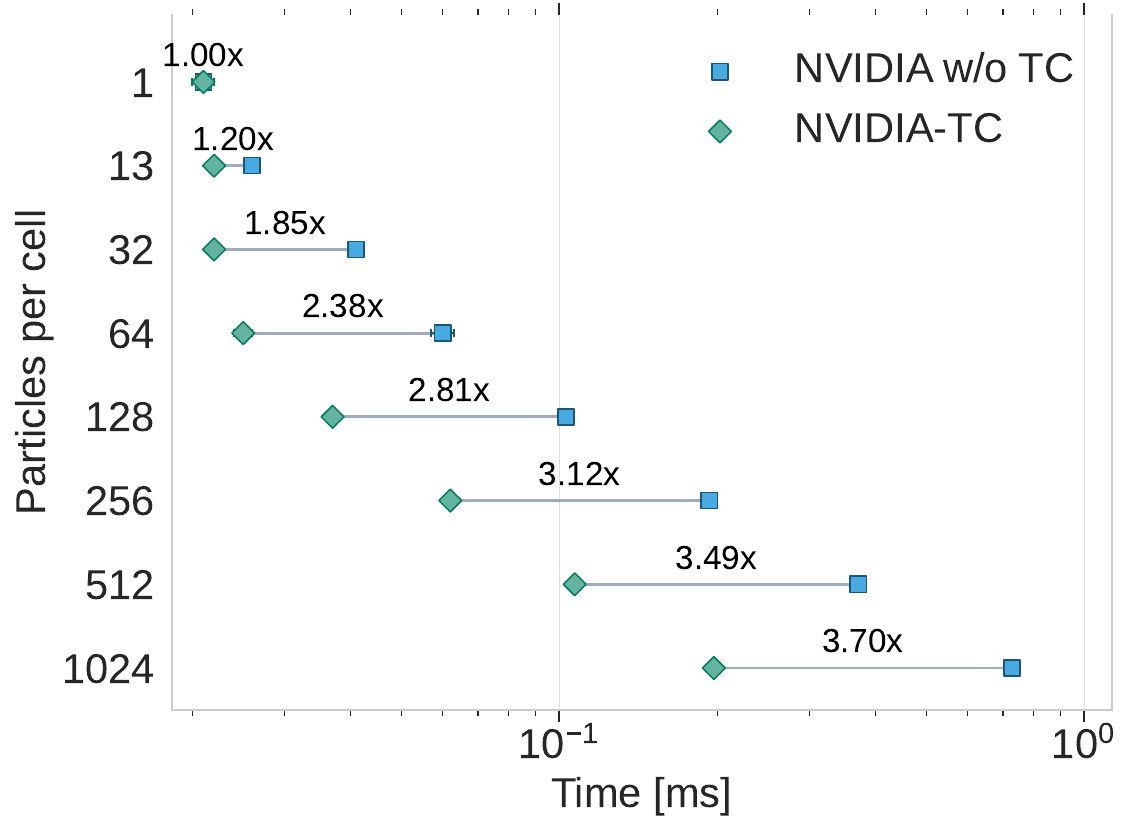}}
    \hspace{1pt}
    \subfloat[]{\includegraphics[width=0.48\linewidth]{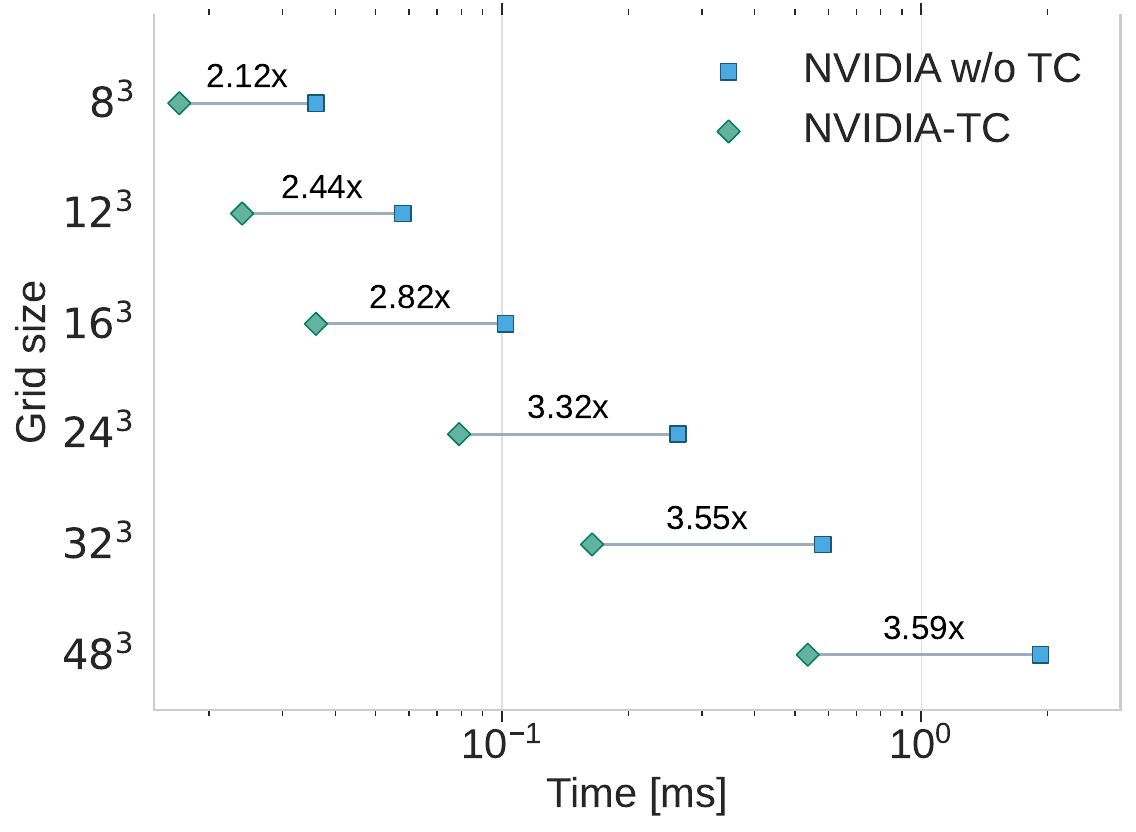}}
    \caption{Tensor cores kernel speedup with respect to a conventional GPU kernel for calculating the scalar mass matrix with CIC interpolation in FP64 precision. a) Performance varying the number of particles per cell, with a fixed grid of $16\times 16\times16$ cells; b) Performance varying the grid size with a fixed number of 128 particles per cell.}
    \label{fig_cic_scalarMM}
\end{figure}

\begin{figure}[t!]
    \centering
    \subfloat[]{
    \includegraphics[width=0.48\linewidth]{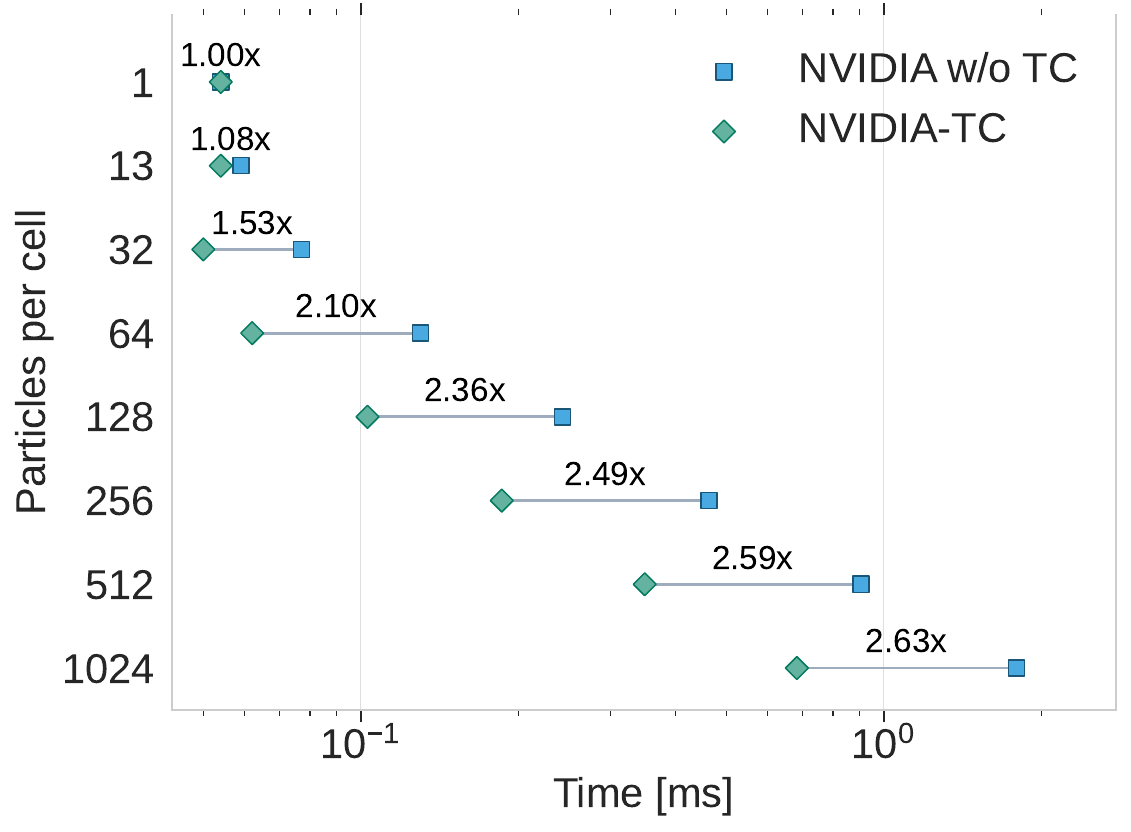}}
    \hspace{1pt}
    \subfloat[]{\includegraphics[width=0.48\linewidth]{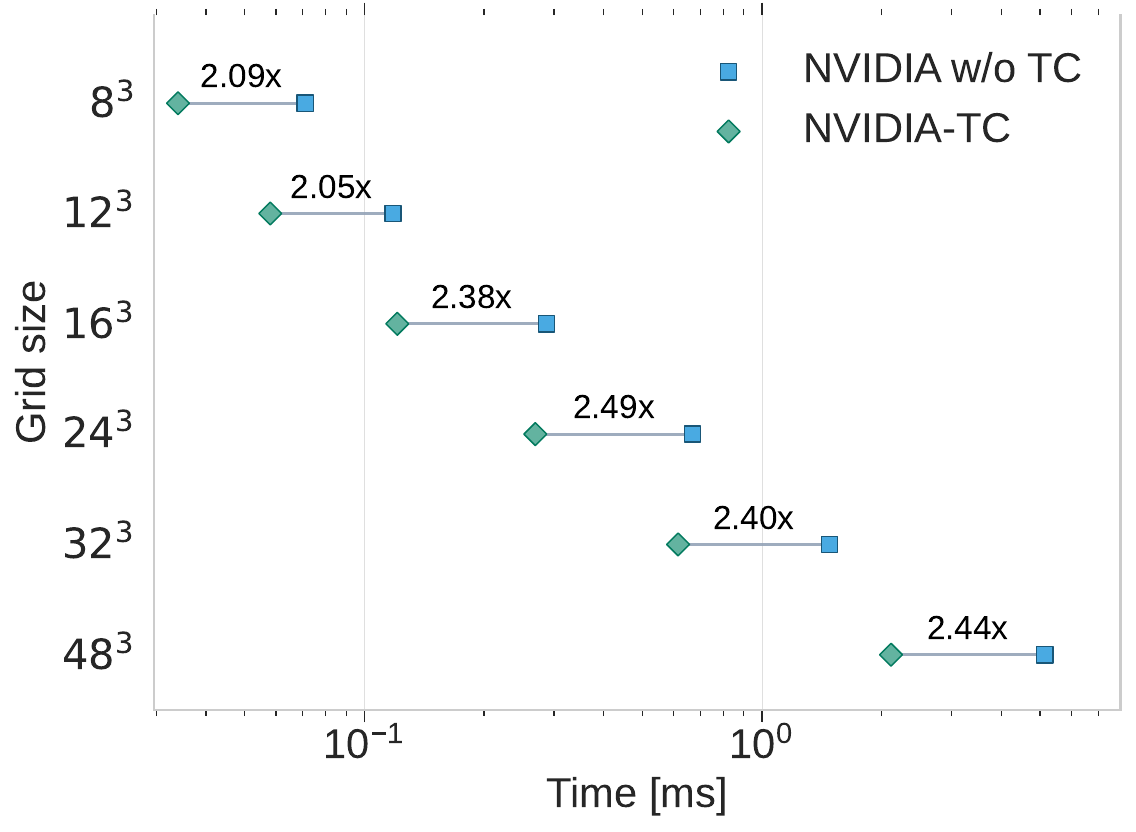}}
    \caption{Tensor cores kernel speedup with respect to a conventional GPU kernel for calculating the $3 \times 3$ ECSIM tensorial mass matrix with CIC interpolation in FP64 precision. a) Performance varying the number of particles per cell, with a fixed grid of $16\times 16\times16$ cells; b) Performance varying the grid size with a fixed number of 128 particles per cell.}
    \label{fig_cic_tensorMM}
\end{figure}

Figure~\ref{fig_cic_scalarMM} reports the CIC scalar mass matrix results in FP64. The WMMA kernel incurs no penalty even at 1~ppc, and its speedup grows monotonically with particle density: from $1.2\times$ at 13~ppc to $3.7\times$ at 1024~ppc. Larger domains also benefit more, with speedups exceeding $2\times$ across all tested grid sizes. Figure~\ref{fig_cic_tensorMM} shows the corresponding CIC results for the $3\times 3$ ECSIM tensorial mass matrix. The trend is similar, with speedups exceeding $2\times$ above 64~ppc and reaching $2.6\times$ at 1024~ppc. The lower acceleration relative to the scalar case is due to the additional non-MMA work in the tensorial kernel, namely loading magnetic field values and precomputing the rotation tensor $\bm\alpha_p$ (Eq.~\eqref{eq_alpha_matrix}). These stages share the same implementation in both kernels and dilute the tensor-core advantage.


\begin{figure}[t!]
    \centering
    \subfloat[]{
    \includegraphics[width=0.48\linewidth]{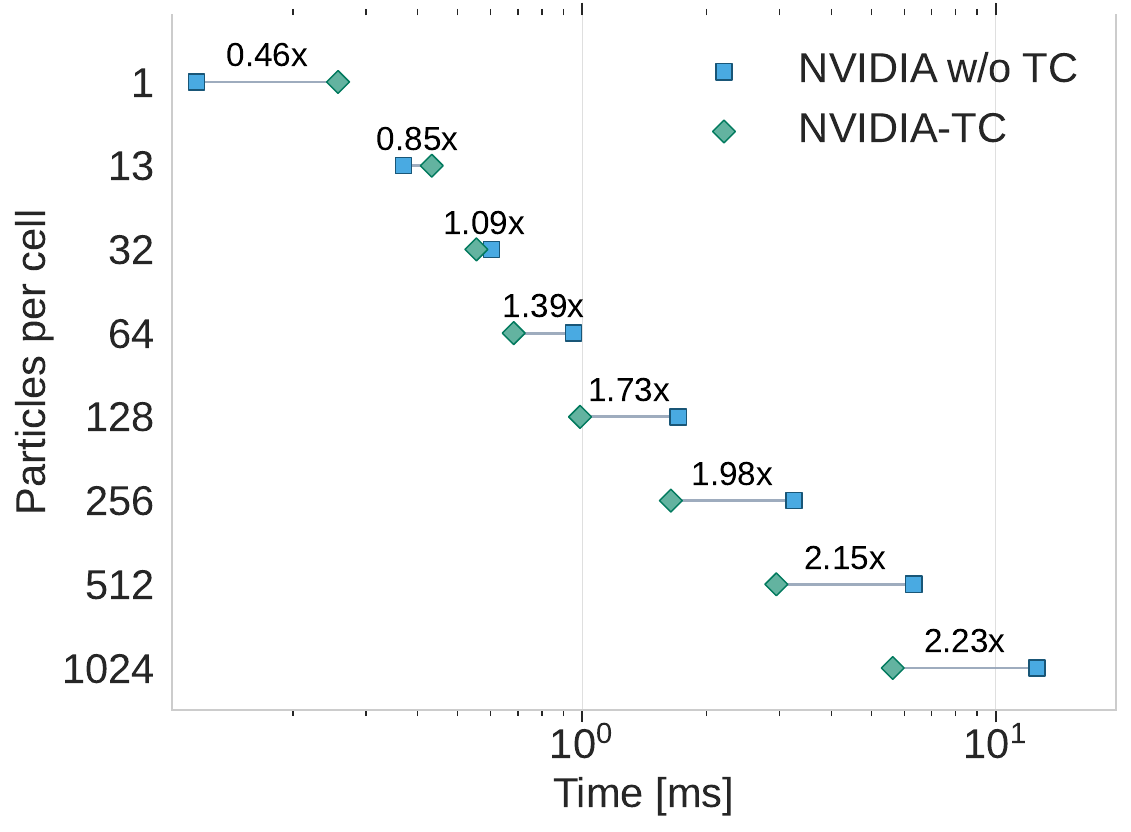}}
    \hspace{1pt}
    \subfloat[]{\includegraphics[width=0.48\linewidth]{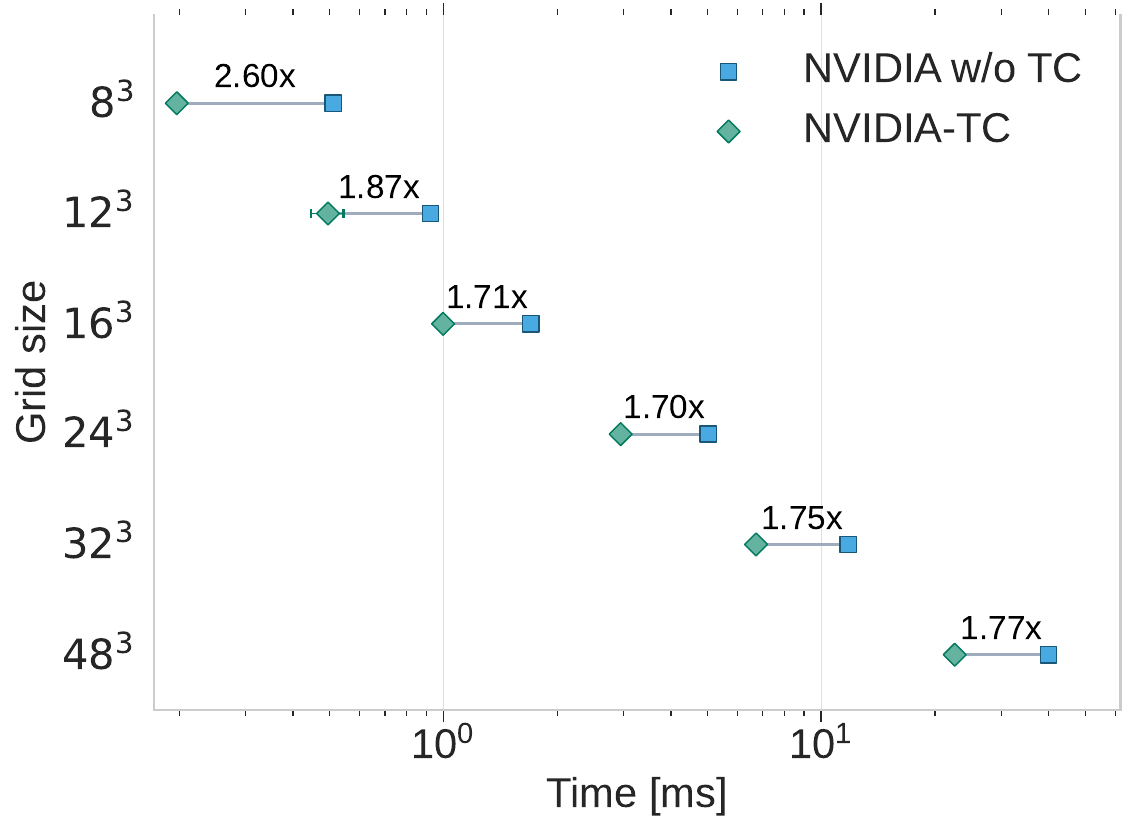}}
    \caption{Tensor cores kernel speedup with respect to a conventional GPU kernel for calculating the scalar mass matrix with TSC interpolation in FP32 precision. a)   Performance varying the number of particles per cell, with a fixed grid of $16\times 16\times16$ cells; b) Performance varying the grid size with a fixed number of 128 particles per cell.}
    \label{fig_tsc_scalarMM}
\end{figure}

\begin{figure}[t!]
    \centering
    \subfloat[]{
    \includegraphics[width=0.48\linewidth]{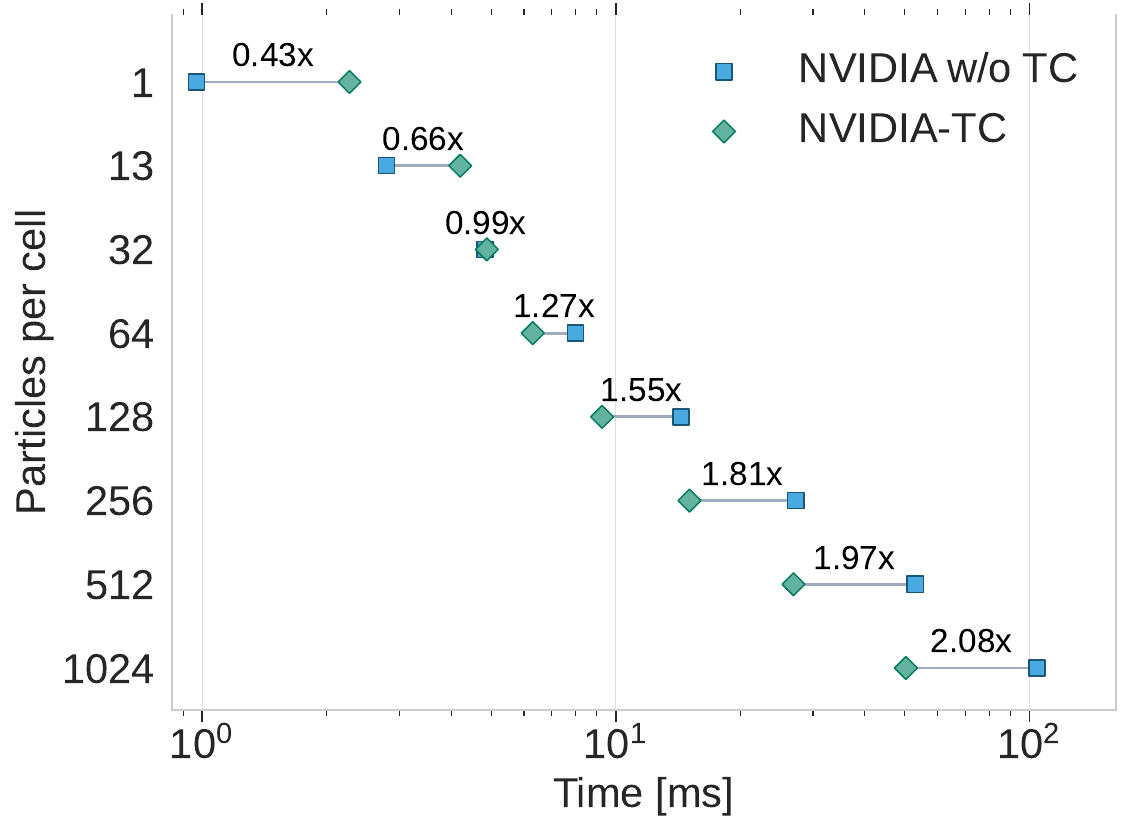}}
    \hspace{1pt}
    \subfloat[]{\includegraphics[width=0.48\linewidth]{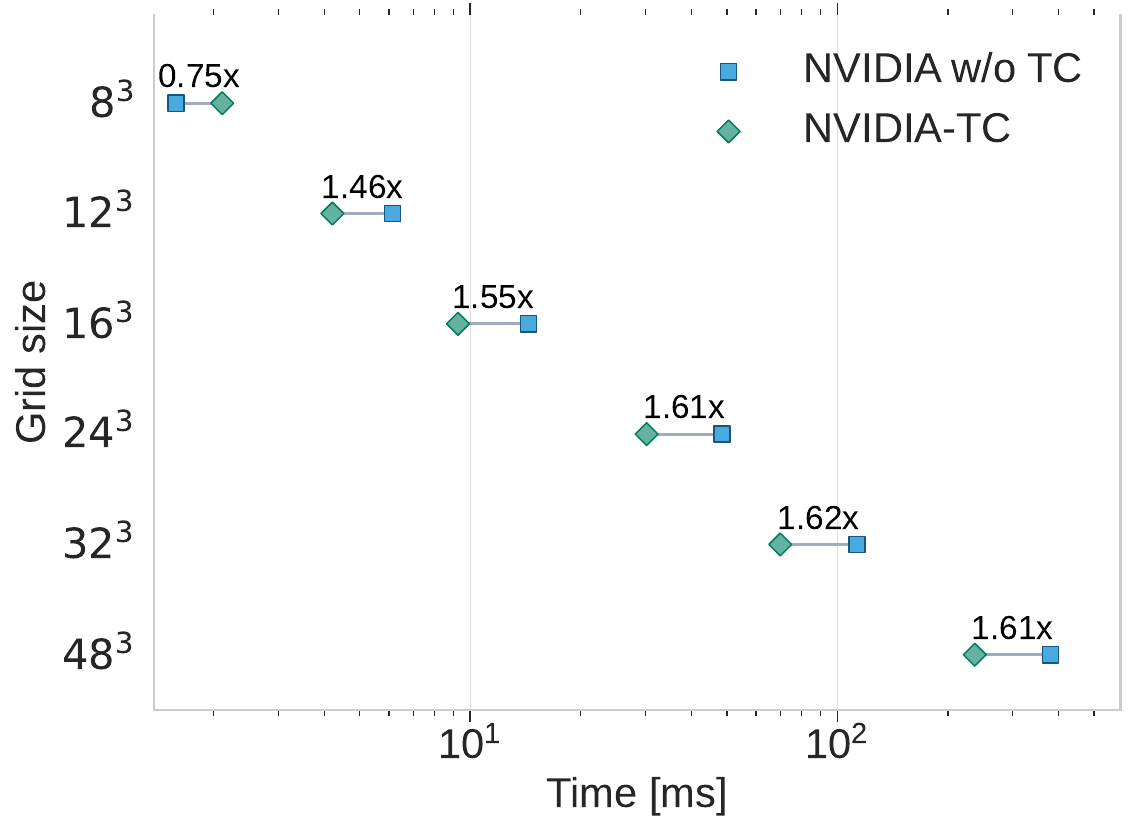}}
    \caption{Tensor cores kernel speedup with respect to a conventional GPU kernel for calculating the $3\times3$ ECSIM tensorial mass matrix with TSC interpolation in FP32 precision. a)   Performance varying the number of particles per cell, with a fixed grid of $16\times 16\times16$ cells; b) Performance varying the grid size with a fixed number of 128 particles per cell.}
    \label{fig_tsc_tensorMM}
\end{figure}

Figures~\ref{fig_tsc_scalarMM} and~\ref{fig_tsc_tensorMM} report the corresponding TSC results in FP32. The overall trend mirrors the CIC case: tensor cores are increasingly beneficial at higher particle densities, with peak speedups at 1024~ppc of $2.2\times$ (scalar) and $\sim 2\times$ (tensorial). The speedups are lower than in the CIC case for two reasons. First, as in the tensorial CIC case, non-MMA pre-computation stages reduce the advantage. Second, particles are sorted by cell but not by support group. We process four of the eight TSC groups per pass, requiring two full passes and thus loading each particle from main memory twice. This overhead is common to both WMMA and non-WMMA kernels but inflates the total cost, reducing the relative gain from tensor cores (peak scalar speedup $2.2\times$ vs.~$3.7\times$ for CIC, tensorial $\sim 2\times$ vs.~$2.6\times$).
With TSC, WMMA shows a slight disadvantage at low particle density ($<32$~ppc) or in small domains. The $2.6\times$ speedup at $8\times 8\times 8$ in Figure~\ref{fig_tsc_scalarMM} b) is an outlier caused by the conventional kernel underperforming, with such a small domain the GPU is heavily underutilized, and the WMMA kernel better hides the control-flow latency of TSC deposition.


\pagebreak
\subsection{End-to-end acceleration of a kinetic plasma simulation}
\label{sec:end_to_end}
\begin{figure}[t]
    \centering
    \includegraphics[width=\linewidth] {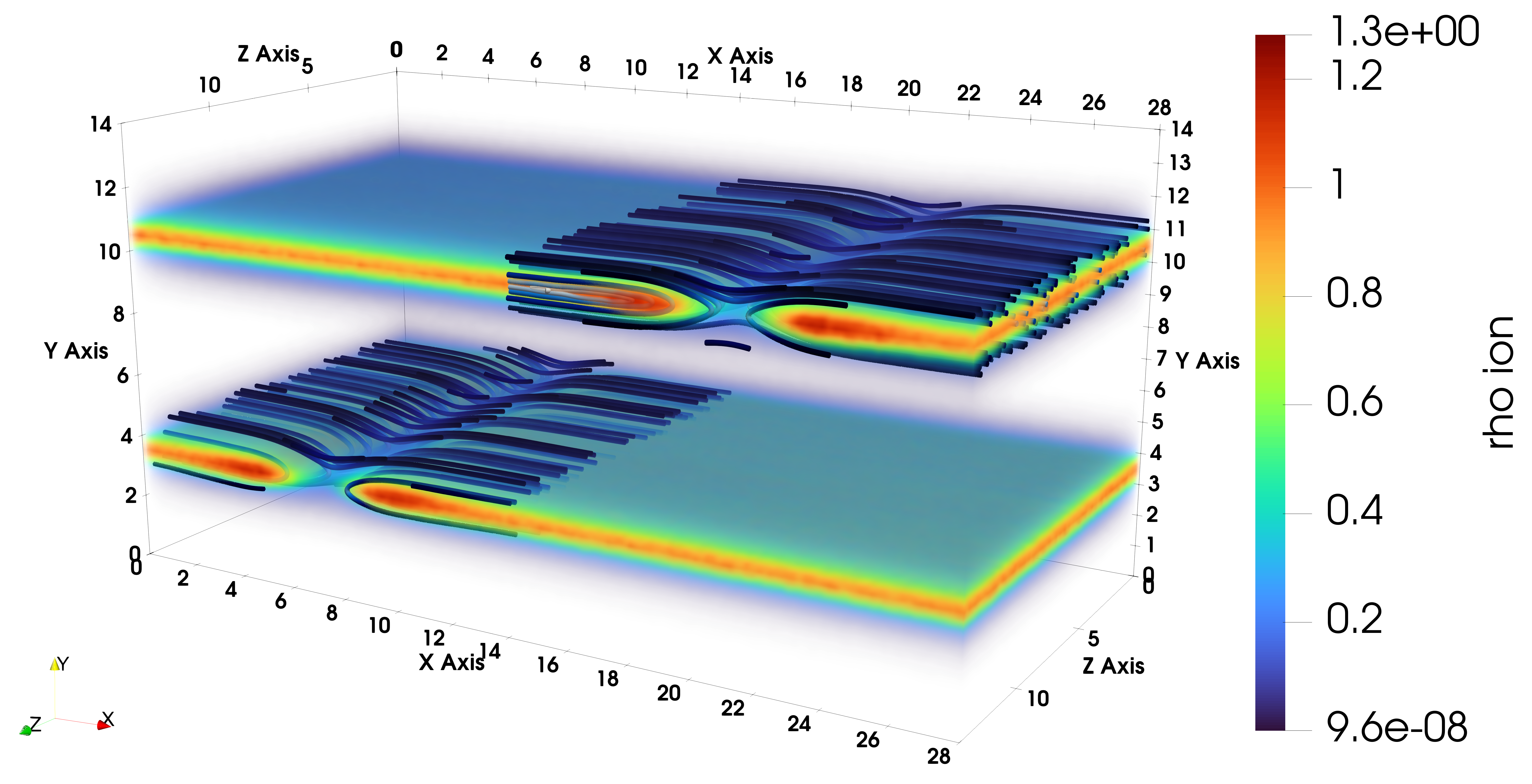} 
    \caption{Charge density of ion species and reconnecting field lines in a 3D double Harris current sheath magnetic reconnection simulation, at $t=600~\omega_{pi}$. The simulation is run accelerating the ECSIM algorithm with tensor cores for the mass matrix assembly.}
    \label{fig_ecsim_gem}
\end{figure}

We assess the end-to-end impact of tensor cores in a production 3D double Harris current sheath magnetic reconnection simulation (Figure~\ref{fig_ecsim_gem}) using the ECSIM algorithm with full periodic boundary conditions. The domain is $28\times 14\times 14\,d_i$ discretized on a $160\times 80 \times 80$ grid, with two species (ions $q/m=1$, electrons $q/m=-256$) at 768~ppc per species ($\sim1.5\times10^9$ particles total).

Since the methodology requires particles sorted by cell, we compare three ECSIM pipeline variants to isolate the contributions of sorting and tensor cores: (i)~unsorted particles with a naive atomic-based mass matrix kernel; (ii)~sorted particles with an optimized conventional GPU kernel; (iii)~sorted particles with WMMA mass matrix assembly. In variants~(ii) and~(iii), sorted particle layout is exploited in all particle-related kernels. The only difference between them is the use of tensor cores in mass matrix deposition.
All three variants use CIC interpolation with FP64 precision and FP64 $(8,8,4)$ tiles, and run with four MPI processes on four NVIDIA A100 GPUs.

\begin{figure}[t!]
    \centering
    \includegraphics[width=0.99\linewidth]{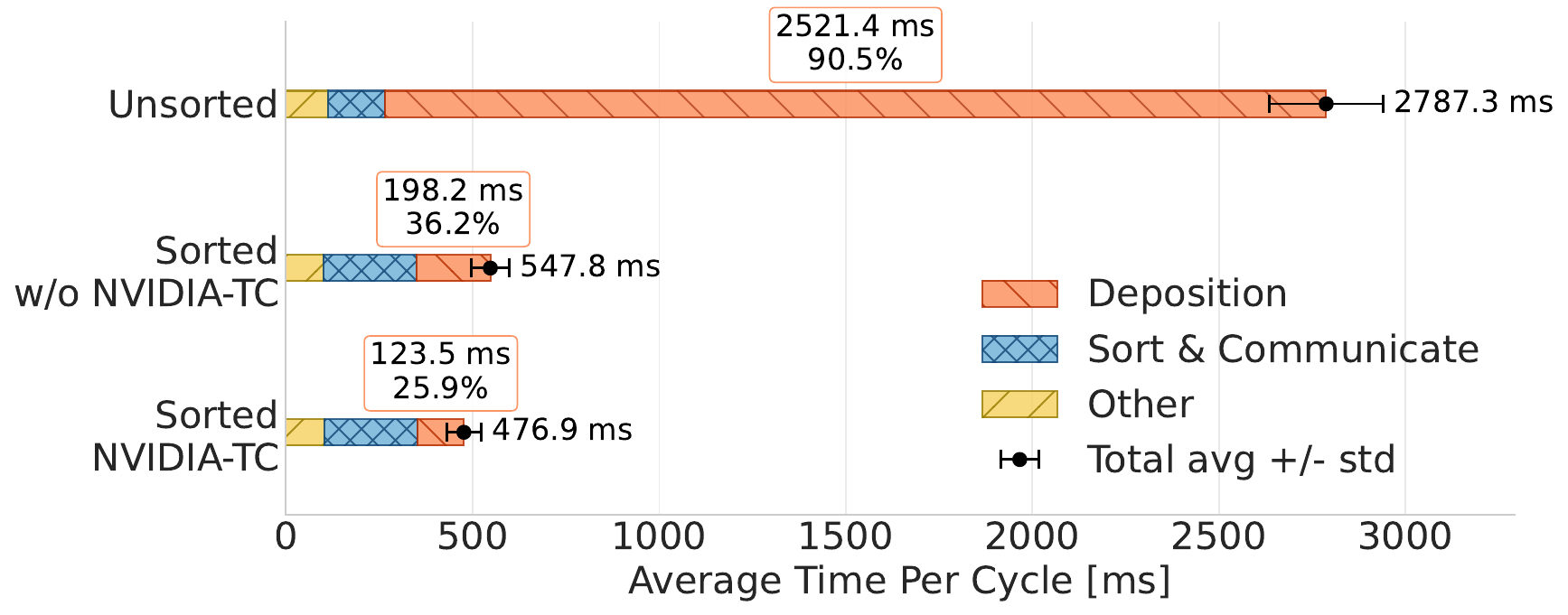}
    \caption{
    Execution time breakdown of a single PIC cycle, averaged over 300 time steps, for a naive ECSIM implementation without particle sorting, a partially optimized implementation with particle sorting without WMMA mass matrix deposition, and a fully optimized implementation that leverages particle sorting and WMMA mass matrix deposition. Colors identify the time spent in the current deposition and mass matrix assembly (orange); particle sorting and communication (blue); particle mover, field solver and diagnostics (yellow). The simulations are run with four MPI processes on 4x NVIDIA A100 GPUs.}
    \label{fig_ecsim_breakdown}
\end{figure}

Figure~\ref{fig_ecsim_breakdown} reports the per-cycle time breakdown, averaged over 300 steps. Because the ECSIM pipeline overlaps MPI communication, host-device transfers, and computation via task-based parallelism, we group the cycle into three non-overlapping stages: \emph{deposition} (mass matrix assembly and moment deposition), \emph{sort \& communicate} (particle sorting and MPI particle exchange), and \emph{other} (field solver, particle mover, diagnostics). I/O is excluded.
In variant~(i), each step takes $\sim2,800$~ms, with deposition accounting for more than $90\%$. Introducing sorting~(ii) reduces \emph{deposition} from $\sim2,500$~ms to $\sim200$~ms at the cost of increasing \emph{sort \& communicate} from $\sim150$~ms to $\sim240$~ms, a strongly favorable trade-off. Tensor cores~(iii) further reduce the \emph{deposition} time by $\sim40\%$ relative to variant~(ii).

Profiling with \texttt{NVIDIA Nsight Systems} shows that the mass matrix kernel alone runs in $\sim24$~ms with WMMA versus $\sim61$~ms without, a $\sim2.5\times$ speedup consistent with the isolation results at 768~ppc (Figure~\ref{fig_cic_tensorMM}). Overall, combining sorting with tensor cores yields a $\sim5.8\times$ end-to-end speedup over the unsorted baseline.

\begin{figure}[t!]
    \centering
    \includegraphics[width=0.58\linewidth]{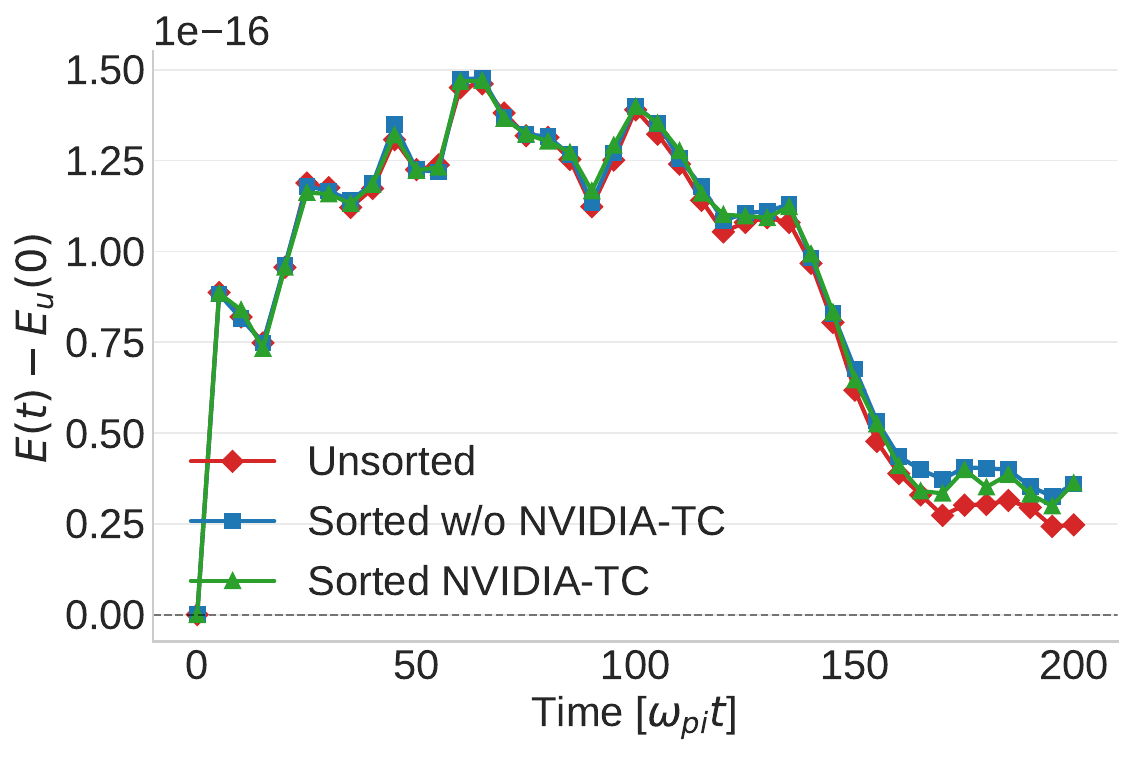}
    \caption{Exact energy conservation to machine precision for the three ECSIM implementations (unsorted, sorted w/o NVIDIA TC, sorted with NVIDIA TC). The plot shows the signed difference between the total system energy $E(t)$ and the total system energy at time $t=0$ measured in the unsorted pipeline, $E_u(0)$.}
    \label{fig_ecsim_energycons}
\end{figure}

To verify physical correctness, we run the WMMA variant for $600~\omega_{pi}$ (Figure~\ref{fig_ecsim_gem}) and compare the total-energy evolution across all three variants over $200~\omega_{pi}$ (Figure~\ref{fig_ecsim_energycons}). The plot shows the signed difference $E(t) - E_u(0)$, where $E_u(0)$ is the initial energy of the unsorted variant. All three implementations preserve energy to machine precision, confirming their physical equivalence.


\pagebreak
\section{Discussion}
\label{sec:discussion}

\paragraph{Kernel-level speedup}
Matrix engines consistently accelerate mass matrix assembly for both CIC and TSC interpolation, and for both scalar and tensorial mass matrices. The magnitude of the speedup depends on the fraction of kernel instructions that can execute on tensor cores:
\begin{itemize}
  \item Lighter kernels (e.g., CIC scalar) benefit most, as particle data are loaded once and no preliminary computation is required, yielding up to $3.7\times$ acceleration.
  \item Heavier kernels (e.g., TSC tensorial) include non-MMA stages, such as magnetic field loading and rotation-tensor precomputation, that reduce the tensor-core advantage; nevertheless, speedups of $\gtrsim 1.5\times$ are still achieved.
\end{itemize}

\paragraph{End-to-end impact}
The $\sim 2.5\times$ kernel-level acceleration observed in isolation is reproduced in the production simulation at 768~ppc. However, the end-to-end speedup depends on the fraction of the PIC cycle spent in mass matrix assembly. In our 3D multi-GPU ECSIM simulation, the optimized pipeline (without tensor cores) already reduces the \emph{deposition} stage to $\sim 36\%$ of the cycle, limiting the tensor-core benefit to $\sim 15\%$ end-to-end. In configurations where deposition dominates, such as 2D simulations without domain decomposition (not reported here), we observed end-to-end speedups of up to $30\%$. Sorting particles by cell is a prerequisite for the MMA-based assembly and incurs a cost of up to $\sim 20\%$ of each cycle, but this is more than compensated by the deposition speedup.

\paragraph{Exactness and precision}
The reformulation is mathematically exact, no approximations are introduced beyond the floating-point rounding inherent in a given tile precision. In our experiments, no rounding error was observed with the FP64 $(8,8,4)$ tile, while a relative error of $\sim 10^{-4}$ was measured with the TF32/FP32 $(16,16,8)$ tile, consistent with the reduced TF32 mantissa.
Mixed-precision workflows are natively supported, since data can be cast on the fly to match the MMA input format. In the production simulation of Section~\ref{sec:end_to_end}, for instance, field values and particle positions are stored in FP32, while velocities, statistical weights, and the mass matrix use FP64.

\paragraph{Tile-support matching and portability}
The approach is most effective when the tile dimensions match the interpolation support. The $(8,8,4)$ tile fits CIC ($N=8$) exactly, and the $(16,16,8)$ tile is well suited to TSC ($N=27$ padded to 32). Higher-order interpolation functions generally benefit from larger tiles; when substantial padding is required, correctness is preserved but throughput is reduced.
The methodology is formulated in terms of a generic MMA abstraction and is directly portable to any platform exposing tile-level MMA instructions. For example, AMD matrix cores offer FP64 $(16,16,4)$ and FP32 $(32,32,2)$ tiles, the latter covers the 27-node TSC support in a single tile without padding.

\section{Conclusion}
\label{sec:conclusion}
In this work, we showed that the mass matrix assembly arising in implicit PIC methods can be reformulated exactly as a tensor contraction that maps onto hardware matrix-multiply-accumulate units. The key observation is that the mass matrix is inherently sparse due to the compact support of the interpolation functions, and the weighted outer-product accumulation over particles decomposes, cell by cell, into a sequence of matrix products whose inner dimension coincides with the contraction dimension of hardware MMA tiles. 

We developed a complete algorithmic framework that includes: i) a cell-local factorization of the global tensor contraction into two-operand matrix products, ii) a particle-batching scheme that partitions particles into groups of size $K_t$ matching the MMA tile inner dimension, and iii) a support-group decomposition that handles the position-dependent stencil placement of higher-order shape functions. The resulting formulation is general with respect to interpolation order, spatial dimension, and the scalar or tensorial nature of the mass matrix, while remaining independent of the specific hardware platform.

We specialized the framework to first-order (CIC) and second-order (TSC) B-spline interpolation in three dimensions and implemented on NVIDIA tensor cores using WMMA intrinsics. Isolation benchmarks on an NVIDIA A100 GPU demonstrated speedups of up to $3.7\times$ for CIC scalar and $2.6\times$ for CIC mass matrices, and up to $2.2\times$ and $\sim 2\times$ for the corresponding TSC cases, with the acceleration being more appreciable at high particle densities. In a production 3D magnetic reconnection simulation using the ECSIM algorithm with CIC interpolation, the tensor-core-accelerated mass matrix kernel achieved a $\sim 2.5\times$ speedup over the optimized conventional GPU kernel, translating into a $\sim 15\%$ reduction of the end-to-end wall-clock time per PIC cycle. 

Importantly, the proposed reformulation is exact and introduces no approximations beyond the floating-point rounding inherent in a given tile precision. The methodology is expressed in terms of a generic MMA abstraction and is therefore directly portable to AMD matrix cores, Intel AMX, Google TPUs, and any future architecture exposing a tile-level MMA interface.\\

More broadly, the same tensor-contraction approach extends to any particle-to-grid scatter operation, where one MMA operand encodes the deposited quantities and the other encodes the interpolation weights. The effectiveness depends on tile occupancy, standard charge and current deposition (four quantities per particle in 3D) would leave an $(8,8,4)$ tile half-occupied, whereas algorithms that scatter higher-order moments, such as the Implicit Moment Method~\cite{brackbill1982_imm, markidis2010_ipic3d,markidis2014fluid} (ten quantities per particle) or high-order moment diagnostic calculations, can fill the tile more efficiently.

\section*{Acknowledgments}
This work is funded by the European Union. This work has received funding from the European High Performance Computing Joint Undertaking (JU) and Sweden, Finland, Germany, Greece, France, Slovenia, Spain, and the Czech Republic under grant agreement No. 101093261, Plasma-PEPSC.

\bibliographystyle{elsarticle-num}
\bibliography{cas-refs}

\newpage
\appendix

\end{document}